\pdfoutput=1
\documentclass{JINST}

\title{Pion and proton showers in the CALICE scintillator-steel analogue hadron calorimeter}

\author{\centering 
\LARGE\bf The CALICE Collaboration
}

\author{\centering
B.\,Bilki$^1$, 
J.\,Repond, 
L.\,Xia
\\ \it
Argonne National Laboratory,
9700 S.\ Cass Avenue,
Argonne, IL 60439-4815,
USA}

\author{\centering
G.\,Eigen 
\\ \it
University of Bergen, Inst.\, of Physics, Allegaten 55, N-5007 Bergen, Norway
}

\author{\centering 
M.\,A.\,Thomson, 
D.\,R.\,Ward
\\ \it
University of Cambridge, Cavendish Laboratory, J J Thomson Avenue, CB3 0HE, UK
}

\author{\centering 
D.\,Benchekroun, 
A.\,Hoummada, 
Y.\,Khoulaki
\\ \it
Universit\'{e} Hassan II A\"{\i}n Chock, Facult\'{e} des sciences.\, B.P. 5366 Maarif, Casablanca, Morocco
}

\author{\centering 
S.\,Chang, 
A.\,Khan, 
D.\,H.\,Kim, 
D.\,J.\,Kong, 
Y.\,D.\,Oh
\\ \it
Department of Physics, Kyungpook National University, Daegu, 702-701,
Republic of Korea
}

\author{\centering
G.\,C.\,Blazey, 
A.\,Dyshkant, 
K.\,Francis, 
J.\,G.\,R.\,Lima, 
R.\,Salcido, 
V.\,Zutshi
\\ \it
NICADD, Northern  Illinois University,
Department of Physics,
DeKalb, IL 60115,
USA
}

\author{\centering
F.\,Salvatore$^2$ 
\\ \it
Royal Holloway University of London,
Dept. of Physics,
Egham, Surrey TW20 0EX, UK
}

\author{\centering 
K.\,Kawagoe,
Y.\,Miyazaki, 
Y.\,Sudo, 
T.\,Suehara, 
T.\,Tomita, 
H.\,Ueno, 
T.\,Yoshioka
\\ \it
Department of Physics, Kyushu University, Fukuoka 812-8581, Japan
}

\author{\centering 
J.\,Apostolakis, 
D.\,Dannheim, 
G.\,Folger, 
V.\,Ivantchenko, 
W.\,Klempt, 
A.\,-I.\,Lucaci-Timoce, 
A.\,Ribon, 
D.\,Schlatter,
E.\,Sicking, 
V.\,Uzhinskiy
\\ \it 
CERN, 1211 Gen\`{e}ve 23, Switzerland
}

\author{\centering 
J.\,Giraud, 
D.\,Grondin, 
J.\,-Y.\,Hostachy, 
L.\,Morin 
\\ \it
Laboratoire de Physique Subatomique et de Cosmologie - Universit\'{e}
Grenoble-Alpes, CNRS/IN2P3, Grenoble, France
}

\author{\centering 
E.\,Brianne,
U.\,Cornett,
D.\,David,
A.\,Ebrahimi,
G.\,Falley,
K.\,Gadow,
P.\,G\"{o}ttlicher,
C.\,G\"{u}nter,
O.\,Hartbrich,
B.\,Hermberg,
S.\,Karstensen,
F.\,Krivan,
K.\,Kr\"{u}ger,
S.\,Lu,
B.\,Lutz,
S.\,Morozov$^3$, 
V.\,Morgunov$^4$, 
C.\,Neub\"user,
M.\,Reinecke,
F.\,Sefkow,
P.\,Smirnov,
H.L.\,Tran
\\ \it
DESY, Notkestrasse 85,
D-22603 Hamburg, Germany
}

\author{\centering
P.\,Buhmann,  
E.\,Garutti, 
S.\,Laurien, 
M.\,Matysek,    
M.\,Ramilli
\\ \it
Univ. Hamburg,
Physics Department,
Institut f\"ur Experimentalphysik,
Luruper Chaussee 149,
22761 Hamburg, Germany
}

\author{\centering 
 K.\,Briggl, 
 P.\,Eckert,  
 T.\,Harion, 
 Y.\,Munwes, 
 H.\,-Ch.\,Schultz-Coulon, 
 W.\,Shen, 
 R.\,Stamen
\\ \it
 University of Heidelberg, Fakultat f\"ur Physik und Astronomie,
Albert Uberle Str. 3-5 2.OG Ost,
D-69120 Heidelberg, Germany
}

\author{\centering
E.\,Norbeck$^5$,  
D.\,Northacker,
Y.\,Onel
\\ \it
University of Iowa, Dept. of Physics and Astronomy,
203 Van Allen Hall, Iowa City, IA 52242-1479, USA
}

\author{\centering 
B.\,van\,Doren,
G.\,W.\,Wilson
\\ \it
University of Kansas, Department of Physics and Astronomy,
Malott Hall, 1251 Wescoe Hall Drive, Lawrence, KS 66045-7582, USA
}

\author{\centering 
M.\,Wing$^6$ 
\\ \it
Department of Physics and Astronomy, University College London,
Gower Street,
London WC1E 6BT, UK
}

\author{\centering 
C.\,Combaret, 
L.\,Caponetto,
R.\,Et\'e, 
G.\,Grenier, 
R.\,Han, 
J.C.\,Ianigro,
R.\,Kieffer, 
I.\,Laktineh,  
N.\,Lumb, 
H.\,Mathez, 
L.\,Mirabito,
A.\,Petrukhin,
A.\,Steen
\\ \it
Universit\'{e} de Lyon, Universit\'{e} Lyon 1, 
CNRS/IN2P3, IPNL 4 rue E Fermi 69622,
Villeurbanne CEDEX, France
}

\author{\centering 
J.\,Berenguer~Antequera,
E.\,Calvo~Alamillo, 
M.-C.\,Fouz, 
J.\,Marin,
J.\,Puerta-Pelayo, 
A.\,Verdugo
\\ \it
CIEMAT, Centro de Investigaciones Energeticas, Medioambientales y Tecnologicas, Madrid, Spain 
}

\author{\centering 
F.\,Corriveau
\\ \it
Department of Physics, McGill University,
Ernest Rutherford Physics Bldg.,
3600 University Ave.,
Montr\'{e}al, Quebec,
Canada H3A 2T8
}

\author{\centering 
B.\,Bobchenko$^7$, 
R.\,Chistov$^7$,   
M.\,Chadeeva$^{7\star}$,  
M.\,Danilov$^{7,8}$, 
A.\,Drutskoy$^7$, 
A.\,Epifantsev, 
O.\,Markin$^7$,  
D.\,Mironov$^{7,8}$,  
R.\,Mizuk$^7$, 
E.\,Novikov, 
V.\,Rusinov$^7$,  
E.\,Tarkovsky$^7$  
\\ \it
Institute of Theoretical and Experimental Physics, B. Cheremushkinskaya ul. 25,
RU-117218 Moscow, Russia
}

\author{\centering 
D.\,Besson, 
P.\,Buzhan, 
A.\,Ilyin, 
E.\,Popova 
\\ \it
National Research Nuclear University 
MEPhI (Moscow Engineering Physics Institute)
31, Kashirskoye shosse,
115409 Moscow, Russia
}

\author{\centering 
M.\,Gabriel, 
C.\,Kiesling,
N.\,van\,der\,Kolk, 
F.\,Simon, 
C.\,Soldner, 
M.\,Szalay, 
M.\,Tesar, 
L.\,Weuste
\\ \it
Max Planck Inst. f\"ur Physik,
F\"ohringer Ring 6,
D-80805 Munich, Germany
}

\author{\centering 
M.\,S.\,Amjad$^9$, 
J.\,Bonis, 
S.\,Callier$^{10}$, 
S.\,Conforti di Lorenzo, 
P.\,Cornebise, 
F.\,Dulucq$^{10}$,  
J.\,Fleury$^{10}$,  
T.\,Frisson$^{11}$, 
G.\,Martin-Chassard$^{10}$, 
R.\,P\"oschl, 
L.\,Raux$^{10}$,
F.\,Richard, 
J.\,Rou\"en\'e,  
N.\,Seguin-Moreau$^{10}$,
Ch.\,de la Taille$^{10}$ 
\\ \it
Laboratoire de L'acc\'elerateur Lin\'eaire,
Centre d'Orsay, Universit\'e de Paris-Sud XI,
BP 34, B\^atiment 200,
F-91898 Orsay CEDEX, France
}

\author{\centering 
M.\,Anduze,
V.\,Boudry, 
J-C.\,Brient, 
C.\,Clerc, 
R.\,Cornat,
M.\,Frotin,
F.\,Gastaldi, 
A.\,Matthieu,   
P.\,Mora de Freitas, 
G.\,Musat, 
M.\,Ruan$^{12}$, 
H.\,Videau
\\ \it
 Laboratoire Leprince-Ringuet (LLR)  -- \'{E}cole Polytechnique,
 CNRS/IN2P3,
 Palaiseau, F-91128 France
}

\author{\centering 
J.\,Zacek 
\\ \it
Charles University, Institute of Particle \& Nuclear Physics,
V Holesovickach 2,
CZ-18000 Prague 8, Czech Republic  
}

\author{\centering 
J.\,Cvach, 
P.\,Gallus, 
M.\,Havranek, 
M.\,Janata, 
J.\,Kvasnicka, 
D.\,Lednicky, 
M.\,Marcisovsky, 
I.\,Polak, 
J.\,Popule, 
L.\,Tomasek, 
M.\,Tomasek,
P.\,Sicho, 
J.\,Smolik, 
V.\,Vrba, 
J.\,Zalesak 
\\ \it
Institute of Physics, Academy of Sciences of the Czech Republic, Na Slovance 2,
CZ-18221 Prague 8, Czech Republic
}

\author{\centering              
D.\,Jeans 
\\ \it
Department of Physics, Graduate School of Science, The University of
Tokyo, 7-3-1 Hongo, Bunkyo-ku, Tokyo 113-0033, Japan
}

\author{{\centering 
S.\,Weber
\\ \it
Bergische Universit\"{a}t Wuppertal
Fachbereich 8 Physik,
Gaussstrasse 20,
D-42097 Wuppertal, Germany
}

\it
$^\star$ Corresponding author\newline
E-mail: \email{marina@itep.ru}
}

\author{  \\
\llap{$^1$}Also at University of Iowa.\\
\llap{$^2$}Now at University of Sussex, Physics and Astronomy Department, Brighton, Sussex, BN1 9QH, UK.\\
\llap{$^3$}Now at Institute for Nuclear Research RAS, Moscow, Russia.\\
\llap{$^4$}Also at Institute of Theoretical and Experimental Physics.\\
\llap{$^5$}Deceased.\\
\llap{$^6$}Also at DESY and Univ. Hamburg.\\
\llap{$^7$}Also at National Research Nuclear University MEPhI.\\
\llap{$^8$}Also at Moscow Institute of Physics and Technology MIPT.\\
\llap{$^9$}Now at COMSATS/Pakistan.\\
\llap{$^{10}$}Now at Laboratoire OMEGA -- \'{E}cole Polytechnique-CNRS/IN2P3, Palaiseau, F-91128 France.\\
\llap{$^{11}$}Now at CERN.\\
\llap{$^{12}$}Now at IHEP, Beijing and CERN.\\
}

\abstract{Showers produced by positive hadrons in the highly granular CALICE scintillator-steel analogue hadron calorimeter were studied. The experimental data were collected at CERN and FNAL for single particles with initial momenta from 10 to 80~GeV/$c$. The calorimeter response and resolution and spatial characteristics of shower development for proton- and pion-induced showers for test beam data and simulations using {\sc Geant4} version 9.6 are compared.}


\begin{document}

\section{Introduction}
\label{sec:intro}

A wide range of highly granular calorimeter prototypes has been developed by the CALICE collaboration to test the particle flow concept as well as new technologies for future particle physics experiments. The particle flow approach (PFA) was proposed in order to achieve the jet energy resolution required for future linear collider experiments~\cite{Videau:2001,Morgunov:2002,Thomson:2009}. 
Recently, a PFA was successfully implemented for jet energy reconstruction in the CMS detector~\cite{CMS:PFA-09,CMS:PFA-10} and a further increase of granularity is now considered an option for the CMS calorimeter upgrade in view of the next phase of the LHC, also called High-Luminosity LHC. Besides testing the PFA, highly granular electromagnetic and hadron calorimeter prototypes provide an opportunity to test Monte Carlo models with unprecedented detail. While the development of electromagnetic showers is quite well understood and reproduced by simulations~\cite{emGeant4}, predictions of hadronic shower development are not so precise and there are no hadronic models which demonstrate agreement with data for all types of hadrons over a wide energy range~\cite{Valid:2013}. 

Hadronic showers, produced in a calorimeter after a deep inelastic interaction of an incident hadron with a nucleus, are characterised by a relatively narrow core from the electromagnetic component surrounded by an extended halo. The core is usually formed by electromagnetic cascades initiated by photons from $\pi^{0}$ decays, while charged mesons and baryons dominate in the radial halo and longitudinal tail of the shower. The complicated structure of hadronic showers results in significant fluctuations of their longitudinal and radial sizes as well as the calorimeter energy response~\cite{Wigmans:2000}.

Besides fluctuations of the response, differences in the average response for different types of hadrons have been predicted~\cite{Gabriel:1994} and observed experimentally for pions and protons. 
The response for pions is $\sim$10\% larger than that for protons in the energy range 200 to 375~GeV~\cite{Akchurin:1998}. Such behaviour was confirmed for lower energies (20 to 180~GeV) with the Fe-Scintillator ATLAS Tile calorimeter~\cite{ATLAS:2010} for which the response to pions was $\sim$4\% higher than the response to protons of the same initial energy.

The CALICE scintillator-steel analogue hadron calorimeter (Fe-AHCAL) is the first example of the large-scale application of silicon photomultipliers (SiPM) in the field of high energy physics~\cite{AHCAL:2010cc}. This calorimeter has high longitudinal and transverse granularity and provides an opportunity for detailed study of hadronic shower development. 
Previous studies of pion-induced showers in the Fe-AHCAL include calorimeter response and resolution, as well as comparisons of pion shower profiles and spatial characteristics with simulations using {\sc Geant4} version 9.4~\cite{Valid:2013,AHCAL:2012res}. 

This paper provides, for the first time, a comparison of the properties of the showers initiated by pions and protons and reconstructed using a high-granularity calorimeter in the energy range from 10 to 80~GeV.
The shower parameters extracted from data are compared with simulations using physics lists from {\sc Geant4} version 9.6~\cite{Geant4:2003}.
The experimental setup, event reconstruction and selection procedure, and systematic uncertainties are described in Section~\ref{sec:exdataMC}. Data-MC comparisons of global observables, such as deposited energy, energy resolution, shower radius, and longitudinal centre of gravity, are presented in Section~\ref{sec:global}.

\section{Experimental data and simulations}
\label{sec:exdataMC}

\subsection{Experimental setup}
\label{sec:setup}

The data analysed here using beams of positively charged hadrons were collected at CERN in 2007 and at FNAL in 2009. The CALICE setup at CERN is described in detail in Ref.~\cite{AHCAL:2012res} and comprised the silicon-tungsten electromagnetic calorimeter (Si-W ECAL), the Fe-AHCAL, and the scintillator-steel tail catcher and muon tracker (TCMT). Beams of positive hadrons in the momentum range from 30 to 80~GeV were delivered with the CERN SPS H6 beam line. The CALICE setup during the test beam campaign at FNAL is described in detail in Ref.~\cite{FeegeDis:2011} and comprised the Fe-AHCAL and TCMT, without any electromagnetic calorimeter in front of the Fe-AHCAL. The data at FNAL were collected for positive hadrons with initial momenta of 10 and 15~GeV from the MTest beam line. 
Data collected with normal incidence of the beam with respect to the calorimeter front plane are used for the current analysis.

The Si-W ECAL is a highly granular sampling electromagnetic calorimeter~\cite{ECAL:2008} comprised of 30 layers and constructed from three sections with different absorber thicknesses. The transverse size of its active zone is 18$\times$18~cm$^2$. The Si-W ECAL has a depth of one nuclear interaction length at normal incidence and a very fine transverse segmentation equivalent to 1$\times$1~cm$^2$ cells.  

The Fe-AHCAL is a sampling structure of 38 active layers interleaved with absorber plates (21~mm of stainless steel per layer). The full transverse size of the calorimeter is 90$\times$90~cm$^2$. Each active layer is assembled from 5 mm thick scintillator tiles of varied transverse sizes: 3$\times$3~cm$^2$ in the central part, 6$\times$6~cm$^2$ in the surrounding region, and 12$\times$12~cm$^2$ in the peripheral region. Each tile is individually read out by a silicon photomultiplier (SiPM). The longitudinal depth of the Fe-AHCAL is $\sim$5.3 nuclear interaction lengths. The calibration procedure for the Fe-AHCAL is described in Ref.~\cite{AHCAL:2010cc}. The calorimeter was positioned so that the beam struck the calorimeter centre tiles to minimise lateral leakage. 

The TCMT is also a sampling calorimeter with 16 active layers assembled from scintillator strips with SiPM readout~\cite{TCMT:2012}. The first nine layers of the TCMT have 2~cm thick absorber plates and the same sampling fraction as the Fe-AHCAL. The absorber thickness of the second TCMT section is larger by a factor of five than in the first section. The total depth of the TCMT amounts to $\sim$5.5 nuclear interaction lengths. 

The experimental setup, both at CERN and at FNAL, included gaseous \u{C}erenkov counters placed upstream of the calorimeters. For the positive hadron test beam data analysed here, the gas pressure was set between the pion and proton thresholds. The information from the \u{C}erenkov counter was not used in the trigger decision during the data acquisition but was recorded for each event in the data set.
This information was used for offline discrimination between pions and protons on an event-by-event basis.
 
The visible signal in each calorimeter cell is obtained in units of minimum-ionising particle (MIP). Only cells with a signal above 0.5~MIP were considered for further analysis; a cell above threshold is called hit. The spatial position of each calorimeter cell is defined in the right-handed Cartesian coordinate system with the $z$-axis oriented along the beam direction, that is perpendicular to the calorimeter front plane, and the $y$-axis pointing up when looking along the beam direction.

\subsection{Monte Carlo simulations}
\label{sec:sim}

Simulations were done using the physics lists {\small QGSP\_BERT} and {\small FTFP\_BERT} from {\sc Geant4} version~9.6 patch~1~\cite{Geant4:2003,Geant4PL:2006}. The physics list {\small QGSP\_BERT} is widely used for simulation in the LHC experiments and has demonstrated the best agreement with data in earlier versions, for instance in the version 9.2~\cite{Geant4Note:2010}. The {\small QGSP\_BERT} physics list is maintained due to its wide use. The physics list {\small FTFP\_BERT} was significantly improved in version 9.6 and is now recommended for HEP simulations by the {\sc Geant4} collaboration~\cite{Geant4:2013proc}.

The {\small QGSP\_BERT} physics list employs the Bertini cascade model ({\small BERT}) below 9.5~GeV, the quark-gluon string precompound model ({\small QGSP}) above 25~GeV, and the low energy parametrised model ({\small LEP}) in the intermediate energy region. The transition regions between models are from 9.5 to 9.9~GeV and from 12 to 25~GeV. 
The {\small FTFP\_BERT} physics list uses the Bertini cascade model for low energies and the Fritiof precompound model ({\small FTFP}) for high energies with a transition region from 4 to 5~GeV.

Separate samples of single pion and single proton events were simulated. The simulated samples were digitised taking into account the SiPM response, light crosstalk between neighbouring scintillator tiles in the same layer, and calorimeter noise extracted from data. The digitisation was validated using the electromagnetic response of the Fe-AHCAL~\cite{AHCAL:2011em}. The test beam profile and its position on the calorimeter front face in each data run were reproduced in simulations. 

\subsection{Spatial observables}
\label{sec:observ}

The longitudinal segmentation of the Fe-AHCAL permits identification of the longitudinal position of the first inelastic interaction of an incoming hadron, which is called the shower start. The shower start is reconstructed on an event-by-event basis as described in Appendix~\ref{app:lambda}. The information about the reconstructed shower start layer is useful for event selection and particle identification which are discussed in Section~\ref{sec:evsel}.

The longitudinal shower development can be studied from either the calorimeter front or the position of the first inelastic interaction with two different observables. The event longitudinal centre of gravity, $Z$, is an energy weighted sum of the longitudinal hit coordinates with respect to the calorimeter front plane and is obtained for each event as

\begin{equation}
Z = \frac{\sum_{i=1}^{N}{e_{i} z_{i}}}{\sum_{i=1}^{N}{e_{i}}},
\label{eq:zcog}
\end{equation}

\noindent where $N$ is the total number of hits in the Fe-AHCAL, $e_i$ is the hit energy, and $z_i$ is the distance from the hit layer to the calorimeter front plane. 

The longitudinal shower depth, $Z0$, represents a longitudinal centre of gravity calculated with respect to the shower start in each event as

\begin{equation}
Z0 = \frac{\sum_{i=1}^{N_{\mathrm{sh}}}{e_{i} \: (z_{i} - z_{\mathrm{start}})}}{\sum_{i=1}^{N_{\mathrm{sh}}}{e_{i}}},
\label{eq:zScog}
\end{equation}

\noindent where $N_{\mathrm{sh}}$ is the number of hits in the Fe-AHCAL from the reconstructed shower start layer and beyond, $e_i$ is the hit energy, $z_i$ is the distance from hit layer to the calorimeter front, and $z_{\mathrm{start}}$ is the distance from the reconstructed shower start layer to the calorimeter front face. In contrast to $Z$, the value $Z0$ does not depend on the shower start position and describes intrinsic longitudinal shower development.

The longitudinal dispersion, $\sigma_{Z0}$, characterises the scattering of shower hits around the longitudinal centre of gravity and is calculated for each event as 
 
\begin{equation}
 \sigma_{Z0} = \sqrt{\frac{\sum_{i=1}^{N_{\mathrm{sh}}}{e_{i} \: (z_{i} - z_{\mathrm{start}})^2}}{\sum_{i=1}^{N_{\mathrm{sh}}}{e_{i}}}  - Z0^2},
\label{eq:sigmaZ0}
\end{equation}

\noindent where $Z0$ is from Eq.~\ref{eq:zScog}.

The radial shower development is characterised by the shower radius, $R$, which is an energy weighted sum of hit radial distances to the shower axis (in the plane perpendicular to the beam direction) and is calculated for each event as 

\begin{equation}
R = \frac{\sum_{i=1}^{N_{\mathrm{sh}}}{e_{i} r_{i}}}{\sum_{i=1}^{N_{\mathrm{sh}}}{e_{i}}},
\label{eq:rmean}
\end{equation}

\noindent where $N_{\mathrm{sh}}$ is the number of hits in the Fe-AHCAL from the shower start layer and beyond, $e_i$ is the hit energy, $r_i = \sqrt{(x_i - x_0)^2 + (y_i - y_0)^2}$ is the distance from the hit with coordinates ($x_i$,$y_i$) to the shower axis with coordinates ($x_0$,$y_0$). The shower axis is defined using the coordinates of the primary track\footnote{The following algorithm is used to find a primary track on an event-by-event basis: a layer by layer search of the single hit candidate per layer is performed in the beam direction, that is, along the normal to the calorimeter front plane, using the nearest neighbour criterion. The search starts from the seed in the first non-empty layer of the calorimeter and ends one layer before the identified shower start. The minimum length of four hits is required for the identified primary track.} in the Si-W ECAL or the coordinates of the event centre of gravity for the data taken without the electromagnetic calorimeter. 
The coordinates of the event centre of gravity are defined as the energy weighted sums of the coordinates of all hits in the given event as $\vec x_{\mathrm{cog}} = (\sum_{i=1}^{N}{e_{i} \: \vec x_{i}})/(\sum_{i=1}^{N}{e_{i}})$.

The radial dispersion, $\sigma_{R}$, characterises the radial scattering of shower hits around the shower radius and is calculated for each event as

\begin{equation}
 \sigma_{R} = \sqrt{\frac{\sum_{i=1}^{N_{\mathrm{sh}}}{e_{i} \: r_{i}^2}}{\sum_{i=1}^{N_{\mathrm{sh}}}{e_{i}}} - R^2},
\label{eq:sigmaR0}
\end{equation}

\noindent where $R$ is from Eq.~\ref{eq:rmean}.

\subsection{Event selection}
\label{sec:evsel}

The purpose of the event selection is to obtain a pure sample of single pion or single proton events for the analysis of the showers produced by these particles in the Fe-AHCAL. The dedicated selection criteria, described in this section, reject events with either an incoming muon, an incoming positron, or two incoming particles. In addition, a dedicated selection is applied, which is based on the information about the reconstructed shower start layer and helps to minimise the longitudinal leakage from the Fe-AHCAL. 
Table~\ref{tab:runList} presents for each beam energy the total number of collected events, the fraction of events rejected by the selection criteria, and the final number of selected pion and proton events. After the selection, the proton samples remain contaminated by single pion events. The measurement of the purity of the proton samples is described in detail in Appendix~\ref{app:ppur}. 

\begin{table}
 \caption{The total number of collected events; the fraction of rejected muon, positron, and double particle events; the number of selected pion and proton events for data samples used in the analysis; and the estimated purity of the selected proton samples.}
 \label{tab:runList}
 \begin{center}
  \begin{tabular}{|c|c|c|c|c|c|c|c|c|}
   \hline
    Beam & Total & Fraction & Fraction & Fraction & Number & Number & Purity & with\\
    momen- & number & of $\mu^{+}$ & of double & of e$^{+}$ & of & of & of selected & ECAL \\
    tum & of  &  & particle &  & selected & selected & proton & in \\
    GeV & events & & events &  & $\pi^{+}$ & protons & sample & front \\
   \hline
    10 &  45839 &  3.0\% & 13.2\% & 37.2\% & 5275 &  1239 & 0.74$\pm$0.13 & no \\
    15 &  46323 &  3.8\% & 13.8\% & 19.9\% & 6660 &  2122 & 0.80$\pm$0.09 & no \\
    30 & 192066 & 30.1\% & 0.2\% &  $\ll 1\%$   & 10838 & 7714 & 0.95$\pm$0.01 & yes \\
    40 & 201069 &  4.6\% & 0.3\% &  $\ll 1\%$   & 20936 & 4799 & 0.84$\pm$0.06 & yes  \\
    50 & 199829 &  4.4\% & 0.3\% &  $\ll 1\%$   & 21151 & 4192 & 0.79$\pm$0.06 & yes  \\
    60 & 208997 &  3.8\% & 0.3\% &  $\ll 1\%$   & 21133 & 5759 & 0.85$\pm$0.05 & yes  \\
    80 & 197062 &  2.8\% & 0.3\% &  $\ll 1\%$   & 16964 & 8545 & 0.83$\pm$0.04 & yes  \\
   \hline
  \end{tabular}
 \end{center}
\end{table}

The muon identification algorithm involves information from all calorimeter sections and is based on the comparison of the energy deposition in the combined ECAL+AHCAL and TCMT. The efficiency of the algorithm was tested with simulations and is better than 99.5\% in the energy range studied. The fraction of identified and rejected muon events does not exceed 5\% for all energies except for the 30~GeV test beam data taken at CERN.

The event is considered to contain more than one incoming particle and is rejected if at least one of the following conditions is satisfied: (a) the reconstructed energy is higher than $E_{\mathrm{beam}} + 2.4\sqrt{E_{\mathrm{beam}}}$, where $E_{\mathrm{beam}}$ is the beam energy in GeV; (b) more than 80~MIP or more than 13 hits are detected in the cells from the first five layers of the Fe-AHCAL which are more than 320~mm far from the Fe-AHCAL centre; (c) parallel incoming tracks are identified before the reconstructed shower start layer. The fraction of such events in the CERN test beam data analysed here was significantly smaller than in the FNAL data samples. A cross-check of the FNAL data with simulations and the CERN data for negative pions of the same energy showed that an additional selection is necessary to reject events with several incoming particles. The following additional constraints were applied to the FNAL data samples: events were rejected with more than 165 and more than 220 hits for 10 and 15~GeV samples, respectively. These selections reject less than 0.2\% of single-hadron events as confirmed by a cross-check with simulations. Table~\ref{tab:runList} shows the fraction of rejected double particle events after all selections applied. 

The pressure in the \u{C}erenkov counter during the data acquisition was set to separate pions and protons, therefore, a signal in the counter can be generated by either pions, muons, or positrons. After the rejection of muon events as described above, the pion samples can remain contaminated by positrons. The admixture of positrons in the selected proton samples is negligible due to the requirement of no signal from the \u{C}erenkov counter in the offline selection of proton events. The rejection of positron events in the data sets analysed here is based on the calorimeter information. For the data samples taken at CERN with the Si-W ECAL in front of the Fe-AHCAL, the requirement of a shower start at the beginning of the Fe-AHCAL fully removes positron contamination, as the longitudinal depth of the Si-W ECAL is 24 radiation lengths.
Another approach was applied to reject positrons in the data taken at FNAL without the electromagnetic calorimeter in front of the Fe-AHCAL. Two characteristics defined in Section~\ref{sec:observ} are used for this selection: event longitudinal centre of gravity, $Z$, and shower radius, $R$, as electromagnetic showers are known to occur close to the front face of the calorimeter and to be more narrow than hadronic showers.
The shower is considered to be induced by a positron if the following conditions are satisfied simultaneously: $R < 37$~mm and $Z < 260$~mm. The rejection efficiency of this criterion was estimated using dedicated positron runs taken at CERN without an electromagnetic calorimeter and was found to be $\sim$96\% at 10~GeV and $\sim$98\% at 15~GeV. The application of this selection to the negative pion samples extracted from CERN data (taken with the Si-W ECAL in front) results in pion rejection of less than 0.8\%. A cross-check with simulations shows that the fraction of proton events rejected by this selection is at least twice smaller than the fraction of pion events. 
 
The \u{C}erenkov counter was used to discriminate between pions and protons in the test beam experiments. As the pressure in the gaseous \u{C}erenkov detector used was set well below the proton threshold, we assume here that the probability of proton contamination in the pion samples is negligible. At the same time, the inefficiency of the \u{C}erenkov counters can result in pion contamination of the proton samples. The estimate of the proton sample purity with respect to pions, $\eta$, based on the independent muon identification, is described in Appendix~\ref{app:ppur}. The estimated purities of the proton samples are listed in Table~\ref{tab:runList} and vary from 74\% to 95\%. 

To minimise leakage into the TCMT, events are required to have a shower start close to the front face of the Fe-AHCAL, that is in the physical layers 2-5 (3-6) in the CERN (FNAL) data analysis. The procedure for shower start identification is described in Appendix~\ref{app:lambda}. 
Events with an identified shower start in the first physical layer of the Fe-AHCAL were excluded from the analysis due to uncertainty associated with the shower start identification. The exclusion of events with a shower start in the first and second Fe-AHCAL layer significantly reduces the fraction of remaining positrons in the data samples taken without an electromagnetic calorimeter. After all selections, the purity of the analysed pion samples with respect to positrons is 0.975$\pm$0.015 and 0.99$\pm$0.01 at 10~GeV and 15~GeV, respectively. 

After requiring a shower start at the beginning of the hadronic calorimeter, the contamination of the selected samples by muons does not exceed 0.1\% for all energies and the admixture of double particle events is less than one percent.
The same procedure of shower start identification and selection by shower start was applied to the data and simulated samples.

\subsection{Systematic uncertainties and biases to observables}
\label{sec:sys}

\paragraph{Systematic uncertainties.} 

The calculation of the reconstructed energy and resolution requires a conversion from MIP response to the GeV energy scale. The conversion coefficient from MIP to GeV for the Fe-AHCAL (electromagnetic calibration) was extracted from dedicated positron runs with a systematic uncertainty of 0.9\%~\cite{AHCAL:2011em}. Other contributions, such as an uncertainty due to the saturation correction of the SiPM response, are discussed in detail in Ref.~\cite{AHCAL:2011em}, they were studied by varying the calibration constants within allowed limits and were found to be negligible. 

The impact of the uncertainty due to the shower start identification on the observables was studied with the simulated samples and was found to be negligible. The shower start uncertainty is assumed to cancel in simulation to data ratios.
This assumption is supported by the fact that the estimates of the nuclear interaction lengths are in agreement between data and simulations as shown in Section~\ref{sec:lam}.

The spatial observables are still affected by leakage due to the limited Fe-AHCAL depth ($\sim$5.3$\lambda^{\mathrm{eff}}_{\mathrm{p}}$) in spite of the applied shower start selection. As shown with simulations, biases from leakage are negligible below 20 GeV and do not exceed a few percent at 80 GeV. The main impact of the leakage is on the longitudinal and radial dispersions. Again, the bias has negligible impact on the comparison of data and simulation.

\paragraph{Correction of contamination bias to observables.} 

The admixture of particle species in the samples introduces a bias to the observables, which can be corrected if the sample purity and parameters of the contaminating sample are available.
Let us consider a measured mean value, $A_{\mathrm{meas}}$, of an observable which is obtained from the contaminated sample with the known purity, $\eta$.  The mean value of the same observable for the contaminating admixture, $A_{\mathrm{cont}}$, is determined independently from a pure sample of contaminating particles. Then the corrected value, $A_{\mathrm{corr}}$, can be calculated as

\begin{equation}
A_{\mathrm{corr}} = A_{\mathrm{meas}}\frac{1}{\eta} + A_{\mathrm{cont}} \left( 1 - \frac{1}{\eta} \right).
\label{eq:bias}
\end{equation}

The uncertainty of the corrected value is calculated using the standard error propagation technique and taking into account the estimated statistical and systematic uncertainties of $A_{\mathrm{meas}}$, $A_{\mathrm{cont}}$, and $\eta$. The purity of the pion and proton samples is quoted in Section~\ref{sec:evsel} and Table~\ref{tab:runList}. 

The positron contamination in the pion samples results in an overestimate of the mean reconstructed energy because the Fe-AHCAL is a non-compensating calorimeter, while the mean longitudinal depth and shower radius are underestimated. The values of the observables for positrons in the Fe-AHCAL were extracted from dedicated positron runs~\cite{AHCAL:2011em}. The most significant correction for bias due to positron contamination is +2.5\% for the mean shower radius of 10~GeV pions. For 15~GeV pion samples, the estimated bias does not exceed 1\% for all observables. 

The admixture of pions in the proton samples results in an overestimate of the reconstructed energy and an underestimate of the longitudinal and radial sizes of the proton showers. The largest biases due to pion contamination are observed for the sample of 10~GeV protons and the corresponding corrections are -4.8\%, +1.3\%, and +2.8\% for the reconstructed energy, longitudinal centre of gravity, and mean shower radius, respectively.
The contamination of the selected samples with muons and double particle events is negligible and does not need correction.

\section{Comparison of observables}
\label{sec:global}

\subsection{Nuclear interaction length}
\label{sec:lam}

\begin{figure}
 \centering
 \includegraphics[width=7.5cm]{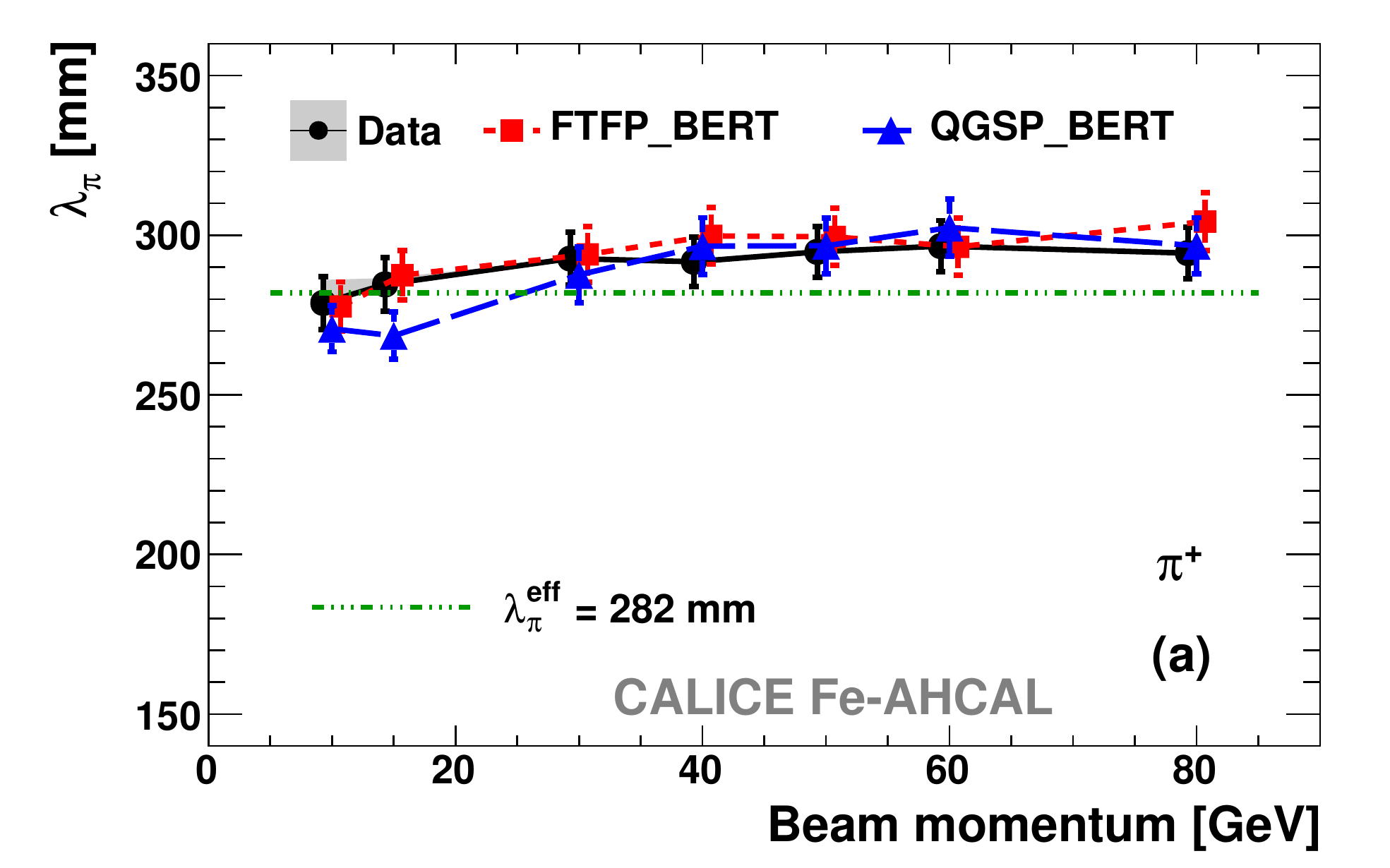}
 \includegraphics[width=7.5cm]{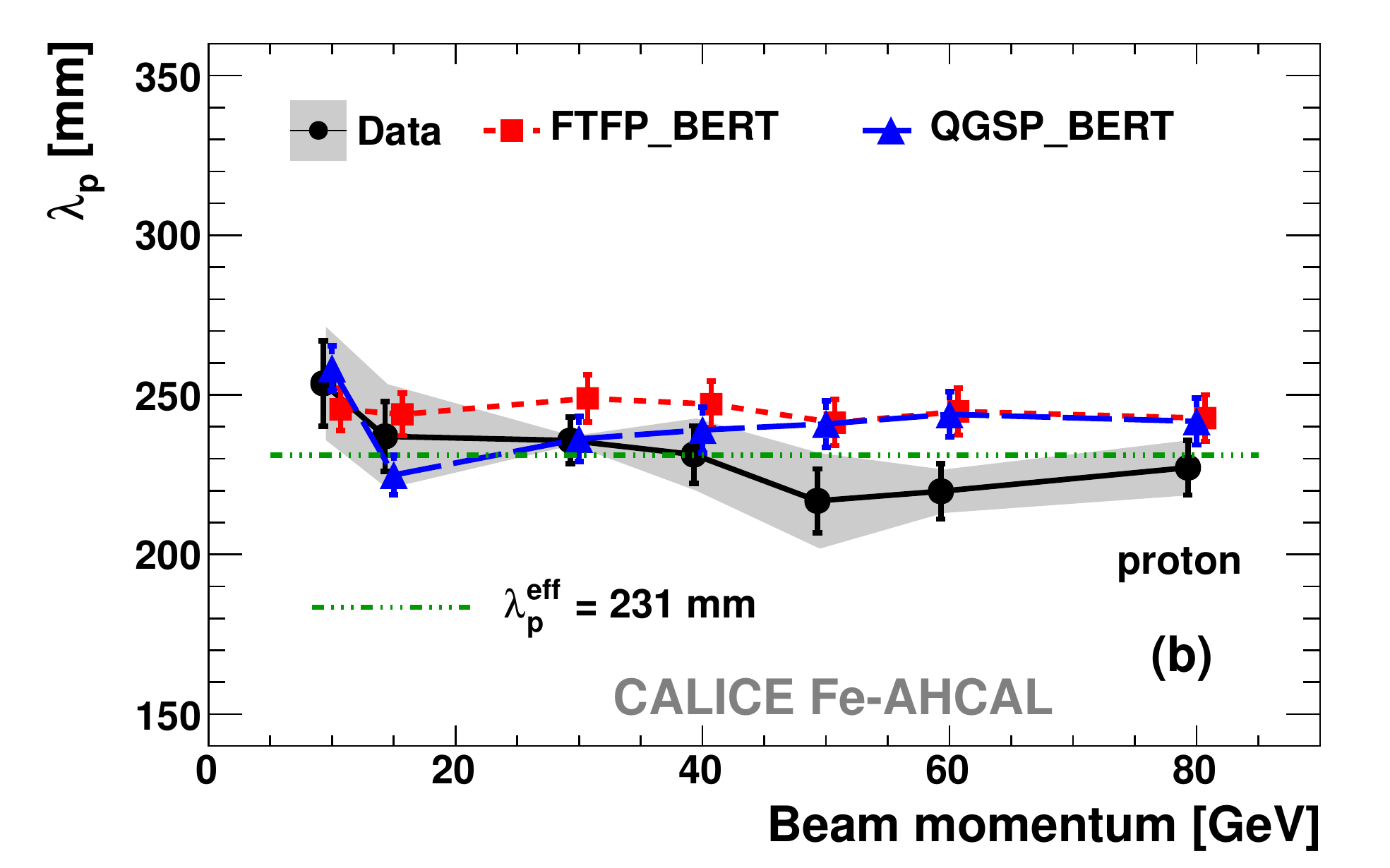}
 \caption{Nuclear interaction length (a) $\lambda_{\pi}$ for pions and (b) $\lambda_{\mathrm{p}}$ for protons in the Fe-AHCAL extracted from reconstructed shower start for data samples (circles, solid lines) and simulations using the {\small FTFP\_BERT} (squares, dotted lines) and {\small QGSP\_BERT} (triangles, dashed lines) physics lists. The error bars represent the uncertainties from the fit. The systematic uncertainties related to sample contamination in the data are shown with grey bands, not visible in the left plot. The dash-dotted lines correspond to the effective nuclear interaction lengths, calculated for the Fe-AHCAL using the material properties data from the PDG tables~\cite{PDG:2012}. }
 \label{fig:lambda}
\end{figure}

The interaction lengths $\lambda_{\pi}$ and $\lambda_{\mathrm{p}}$ for pions and protons were extracted from the distributions of the reconstructed shower start layer as described Appendix~\ref{app:lambda}. The energy dependencies of these estimates are shown in Fig.~\ref{fig:lambda}. The systematic uncertainties for $\lambda_{\mathrm{p}}$ are dominated by pion contamination of the proton samples. 
The interaction length estimated for pion data above 20~GeV is in better agreement with the simulations than that for proton data.

The dash-dotted lines in Fig.~\ref{fig:lambda} show the effective nuclear interaction lengths $\lambda^{\mathrm{eff}}_{\pi}$ and $\lambda^{\mathrm{eff}}_{\mathrm{p}}$, which were calculated for the compound structure of the CALICE Fe-AHCAL using data on material properties from the PDG tables~\cite{PDG:2012}. The variation of the interaction length extracted from data as a function of the pion or proton energy is compatible with a constant with $\frac{\chi^2}{\mathrm{NDF}} < 1.3$. The average value of $\lambda_{\mathrm{p}}$, estimated from the proton data in the energy range studied, is in agreement with $\lambda^{\mathrm{eff}}_{\mathrm{p}}$ within uncertainties. The average value of $\lambda_{\pi}$ is overestimated compared to $\lambda^{\mathrm{eff}}_{\pi}$.

\subsection{Calorimeter response and p/$\pi$ ratio}
\label{sec:resp}

The total energy deposited by a particle is reconstructed at the electromagnetic scale and expressed as

\begin{equation}
E_{\mathrm{event}}= \left\{
\begin{array}{rr}
E_{\mathrm{HCAL}} + E_{\mathrm{TCMT}} & \quad \mathrm{for \; FNAL \; data,}\\ 
E_{\mathrm{ECAL}}^{\mathrm{track}} + E_{\mathrm{HCAL}} + E_{\mathrm{TCMT}} & \quad \mathrm{for \; CERN \; data,}
\end{array}
\right.
\label{eq:reco}
\end{equation}

\noindent where $E_{\mathrm{ECAL}}^{\mathrm{track}}$, $E_{\mathrm{HCAL}}$, and $E_{\mathrm{TCMT}}$ are the energies deposited in the Si-W ECAL, in the Fe-AHCAL, and in the TCMT, respectively. The deposited energies are obtained by multiplying the visible signal in units of MIP by a suitable conversion factor from MIP to GeV for each detector section. 

The energy deposited in the Si-W ECAL is that of a minimum ionising particle, as the events are selected with the shower start at the beginning of the Fe-AHCAL. The obtained conversion factor\footnote{This factor was calculated as a ratio of the mean total energy in units of GeV, deposited in the Si-W ECAL by simulated muons, to the mean visible signal in units of MIP, measured in the Si-W ECAL for muons from the dedicated muon runs.} for a minimum ionising particle in the Si-W ECAL is 0.0030$\pm$0.0002~GeV/MIP. The value of $E_{\mathrm{ECAL}}^{\mathrm{track}}$ for the incident energies from 30~GeV and above accounts for less than 1.4\% of the reconstructed energy and is on average $\sim$0.35~GeV.
The conversion factors for the Fe-AHCAL and the TCMT are obtained from the electromagnetic calibration factor. The electromagnetic calibration factor for the Fe-AHCAL was extracted from dedicated positron runs~\cite{AHCAL:2011em} and is 0.0236$\pm$0.0002~GeV/MIP. Since the first nine TCMT layers are essentially identical to the Fe-AHCAL layers in terms of absorber and active material, the same electromagnetic calibration factor is assumed. For the last seven TCMT layers, this factor is adjusted according to the increased absorber thickness. 

The reconstructed energy distributions were fitted with a Gaussian curve in the interval of $\pm 2$~r.m.s. around the mean value. Hereafter, the parameters of this Gaussian fit at a given beam energy are referred to as the mean reconstructed energy $E_{\mathrm{reco}}$ and resolution $\sigma_{\mathrm{reco}}$. 

\begin{figure}
 \centering
 \includegraphics[width=7.5cm]{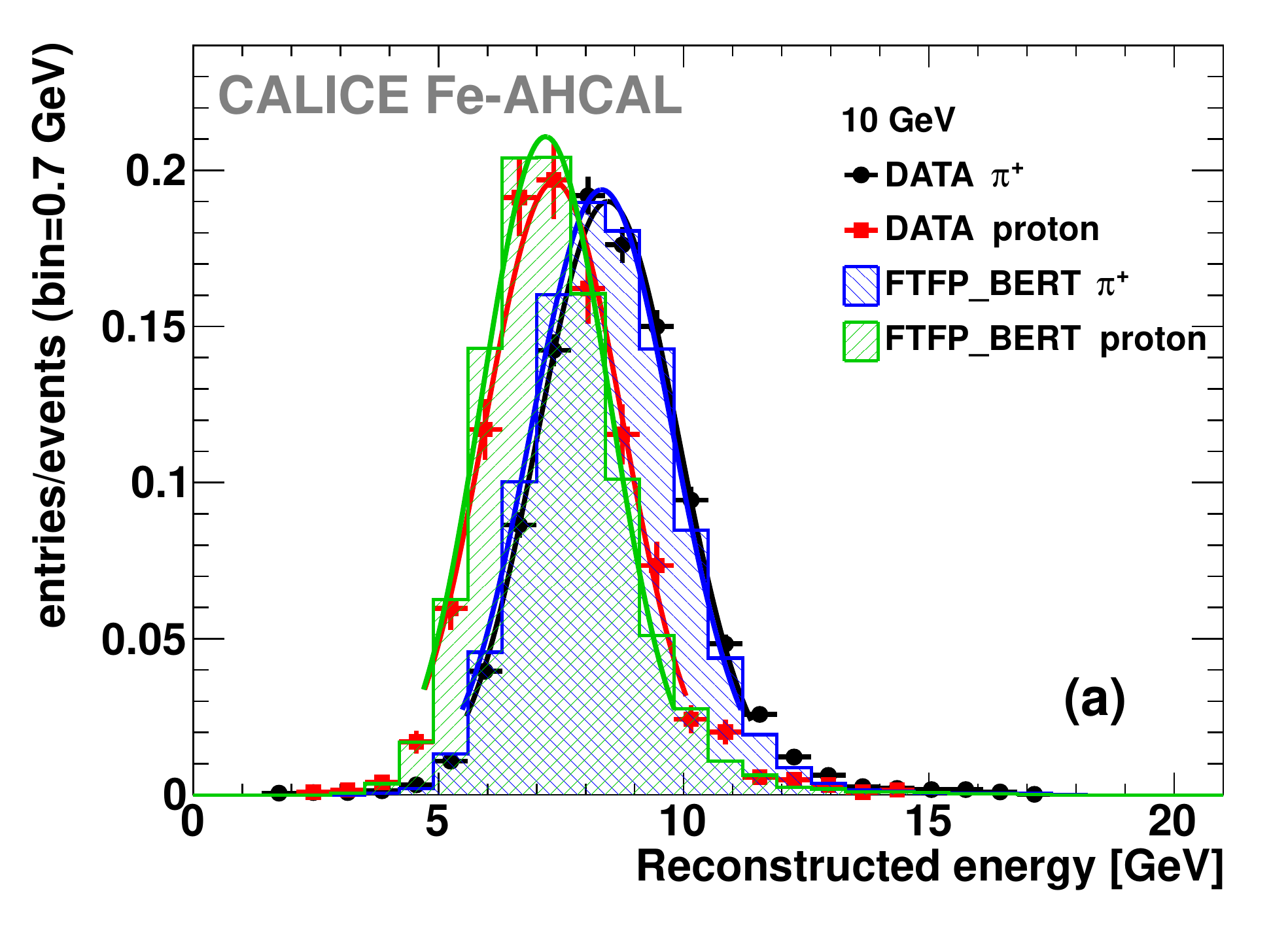}
 \includegraphics[width=7.5cm]{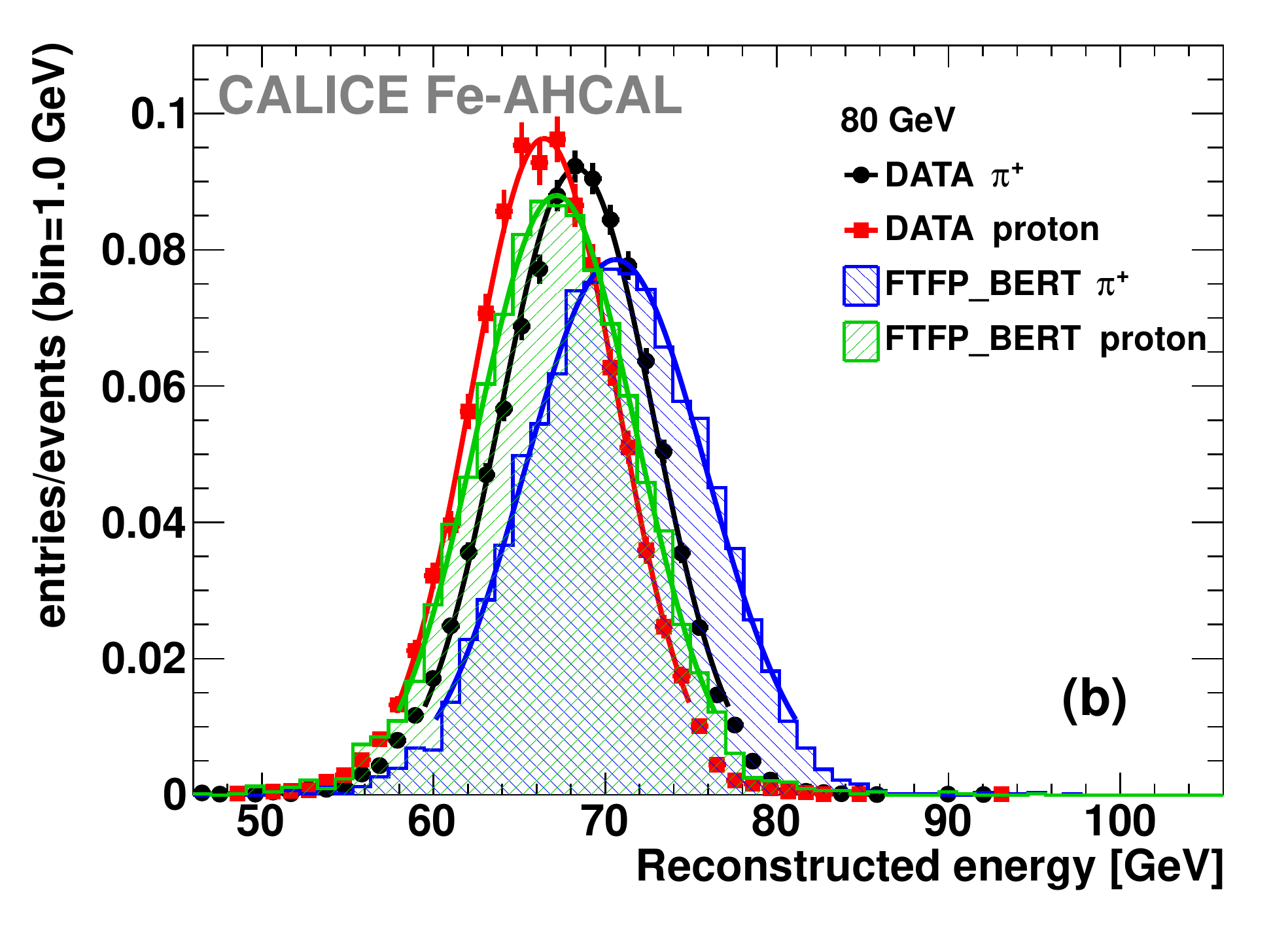}
 \caption{Reconstructed energy distributions for pions and protons with initial energies (a) 10 and (b) 80~GeV for data (points) and simulations using the {\small FTFP\_BERT} physics list (hatched histograms). The solid curves are Gaussian fits to the data and simulations. Error bars show the statistical uncertainties.}
 \label{fig:erecodist}
\end{figure}

Two examples of the reconstructed energy distribution are shown in Fig.~\ref{fig:erecodist} for pions and protons of 10 and 80~GeV together with the predictions of the {\small FTFP\_BERT} physics list. In agreement with the earlier published results~\cite{Gabriel:1994,Akchurin:1998,ATLAS:2010}, the reconstructed energy for protons is lower than that for pions. The relative difference increases with decreasing initial particle energy. This behaviour can largely be explained by baryon number conservation that results in lower probability to produce a leading baryon in the interaction of a pion with a nucleus. Therefore, the measurable energy is different for pions and protons and corresponds to the total particle energy in the case of mesons and to the kinetic energy in the case of baryons

\begin{equation}
E^{\mathrm{proton}}_{\mathrm{available}} = \sqrt{p^{2}_{\mathrm{beam}} + m^{2}_{\mathrm{proton}}} -  m_{\mathrm{proton}},
\label{eq:available}
\end{equation}

\noindent where $p_{\mathrm{beam}}$ is the beam momentum and $m_{\mathrm{proton}}$ is the proton rest mass.

\begin{figure}
 \centering
 \includegraphics[width=7.5cm]{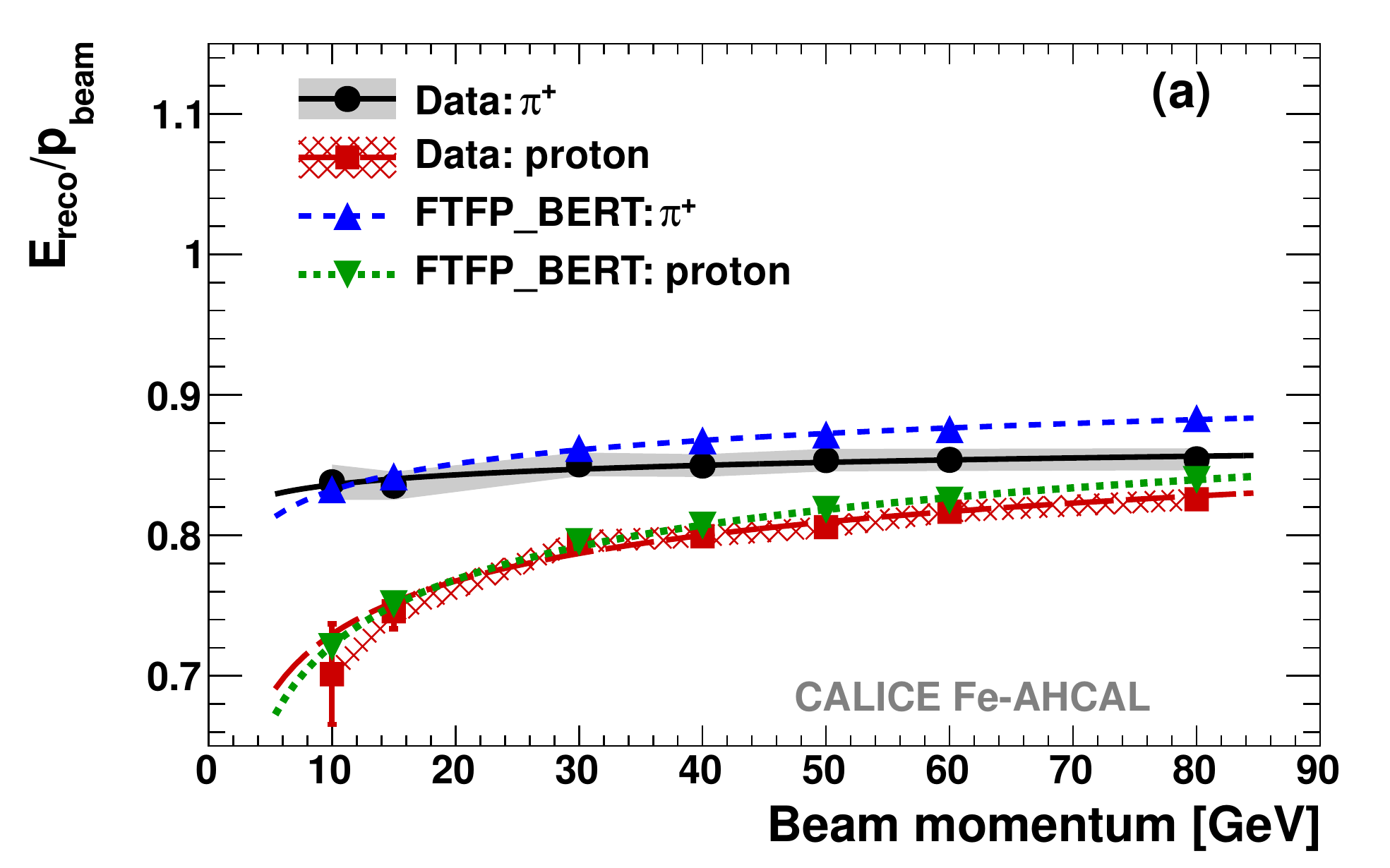}
 \includegraphics[width=7.5cm]{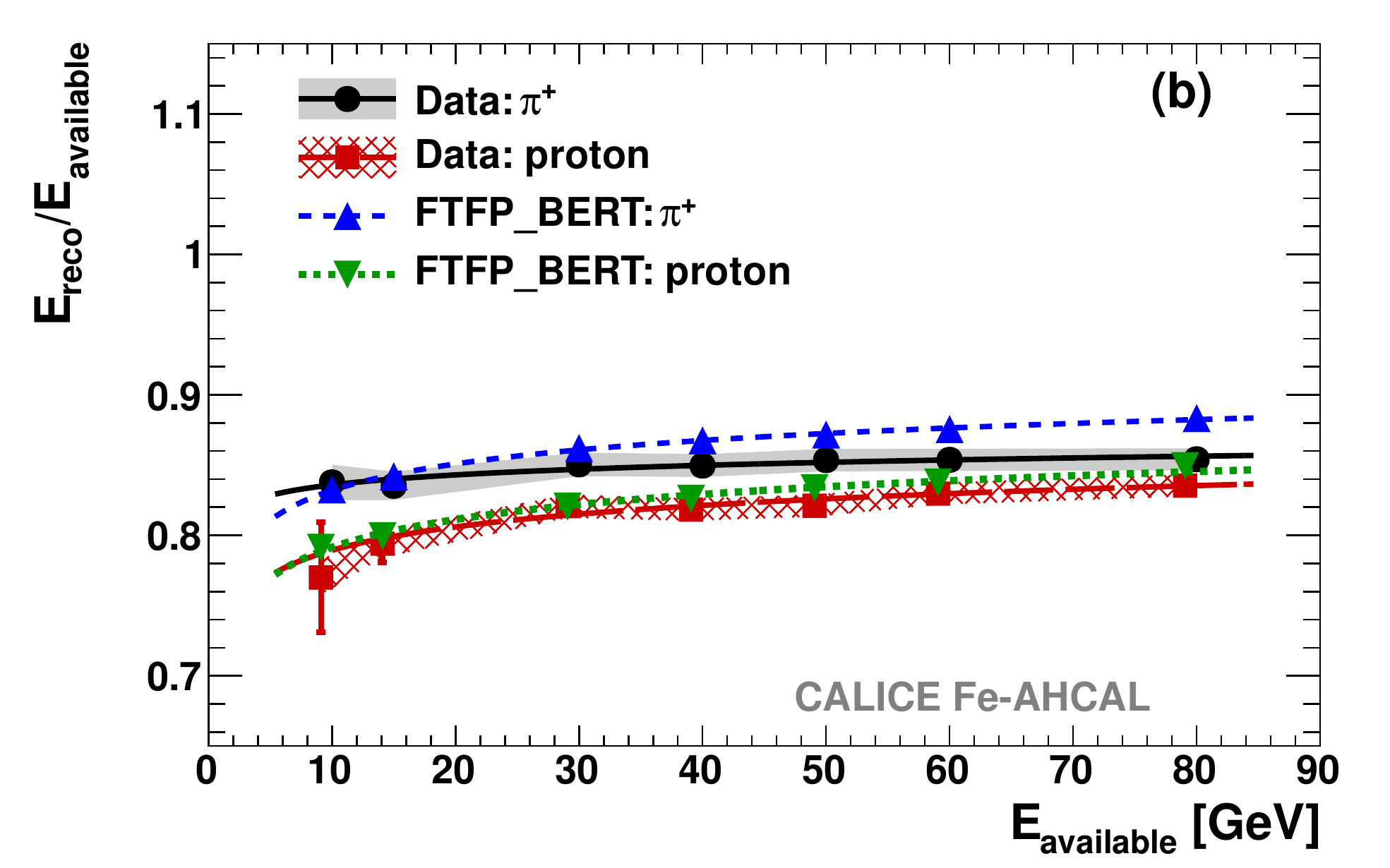}
 \includegraphics[width=7.5cm]{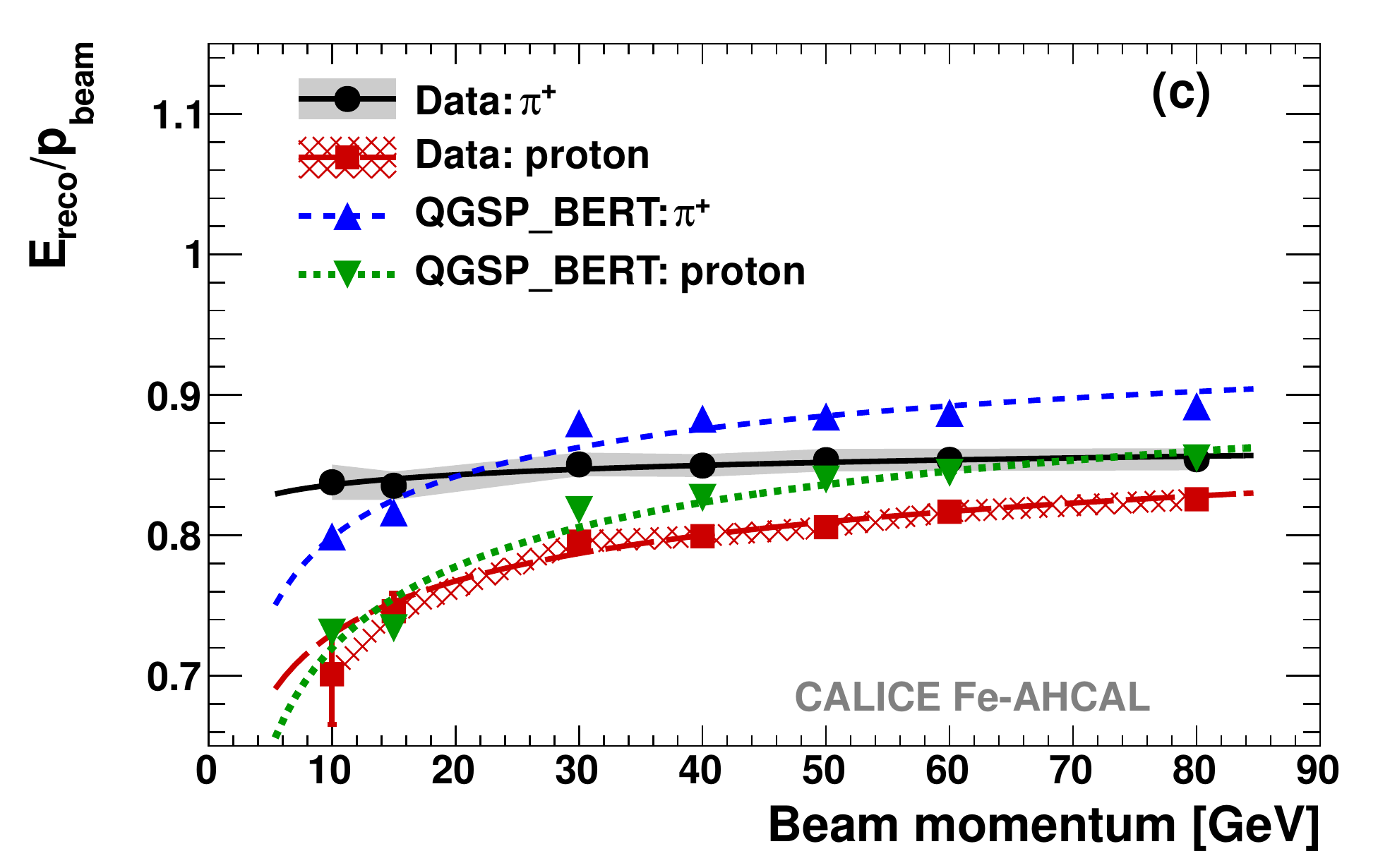}
 \includegraphics[width=7.5cm]{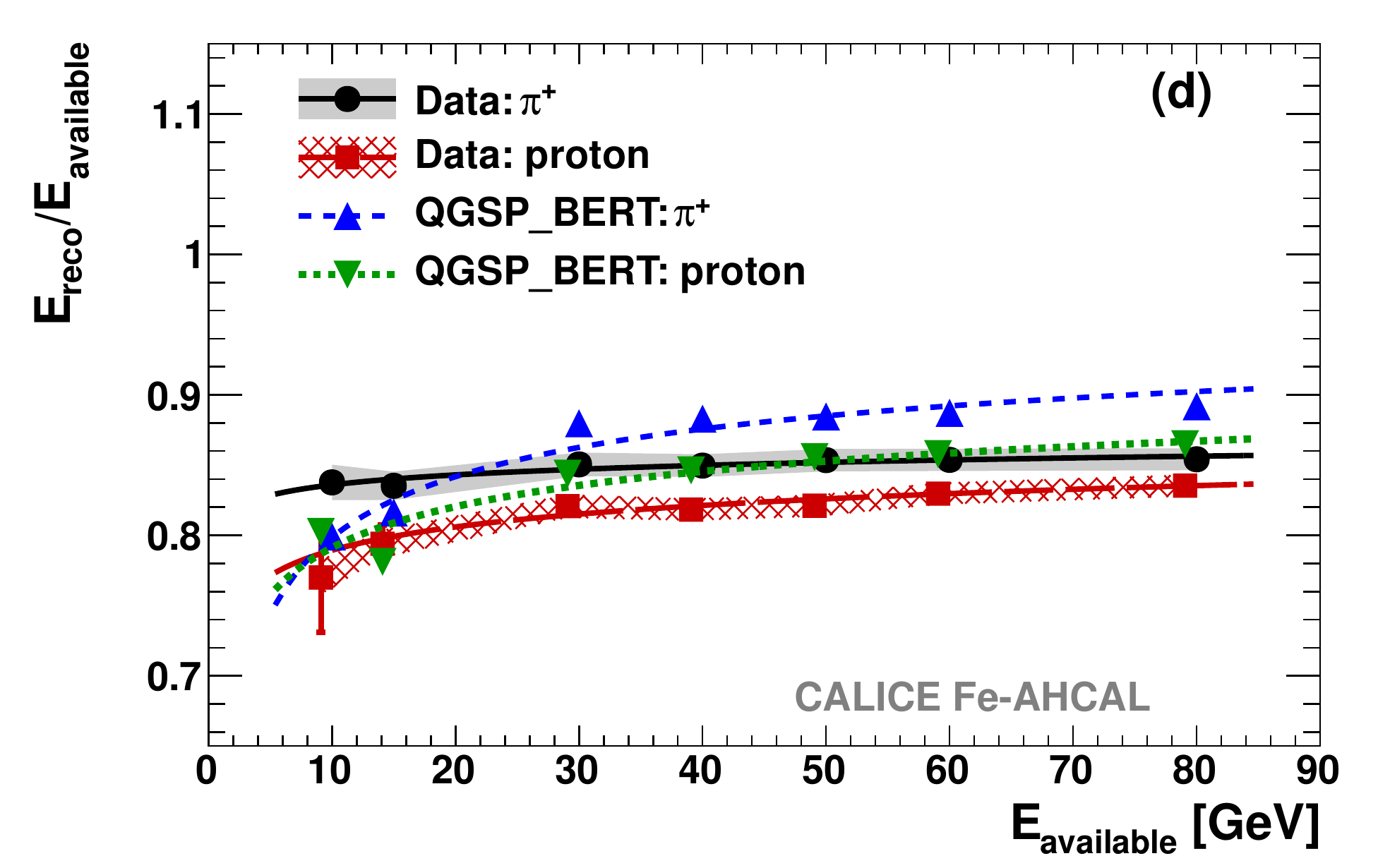}
 \caption{Ratio of the mean reconstructed energy $E_{\mathrm{reco}}$ to (a,c) beam momentum and (b,d) available energy for data and the {\small FTFP\_BERT} (upper row) and {\small QGSP\_BERT} (bottom row) physics lists. The values of $E_{\mathrm{reco}}$ for data are corrected for contamination bias as described in Section~\protect\ref{sec:sys}. Error bars show the statistical uncertainties. The filled and crosshatched bands show the systematic uncertainties for pion and proton data, respectively. The curves correspond to the power-law approximation.}
 \label{fig:linFtfp}
\end{figure}

Figure \ref{fig:linFtfp} shows the ratios of the mean reconstructed energy to beam momentum and to the available energy for data and simulations with the {\small FTFP\_BERT} and {\small QGSP\_BERT} physics lists. The mean reconstructed energy is corrected for the contamination bias as described in Section~\ref{sec:sys}. The difference between positive pion and proton response to the available energy remains at the level of $\sim$5\%, in agreement with the difference observed for the Sc-Fe Tile ATLAS calorimeter~\cite{ATLAS:2010}. The {\small FTFP\_BERT} physics list gives better predictions of the response for pions than {\small QGSP\_BERT} and very good predictions for protons.

The phenomenological interpretation of the observed response behaviour is usually done in terms of the mean electromagnetic fraction, $f_{\mathrm{em}}$, within a hadron-induced shower and the mean responses, $e$ and $h$, to the electromagnetic and hadronic components, respectively. The factor $e$ also defines the electromagnetic scale and can be extracted from electromagnetic calibration. In the frame of such an approach, the mean reconstructed energy can be expressed as

\begin{equation}
E_{\mathrm{reco}} = E_{\mathrm{beam}} (f_{\mathrm{em}} + \frac{h}{e}\:f_{\mathrm{h}}),
\label{eq:respfrac}
\end{equation}

\noindent where $E_{\mathrm{reco}}$ is the mean reconstructed energy of pions ($E_{\pi}$) or protons ($E_{\mathrm{p}}$) measured at the electromagnetic scale, $E_{\mathrm{beam}}$ is the beam energy, and $f_{\mathrm{h}} = 1 - f_{\mathrm{em}}$ is the mean hadronic fraction.
Considerations about the cascade shower development and constraints on $\pi^{0}$ production~\cite{Groom:2007} lead to the assumption of a power-law scaling of the mean hadronic fraction:
 $f_{\mathrm{h}} \approx (E_{\mathrm{beam}}/E_{0})^{m-1}$,
 where $E_{0}$ is the energy at which multiple pion production becomes significant and which is expected to be different for pions and protons. Applying this scaling to Eq.~\ref{eq:respfrac} gives the calorimeter response
 
\begin{equation}
\frac{E_{\mathrm{reco}}}{E_{\mathrm{beam}}}
 = 1 - \left (1 - h/e \right ) \: f_{\mathrm{h}} = 1 - a \: E_{\mathrm{beam}}^{m-1}, \quad a = (1 - h/e) \: E_{0}^{1-m},
\label{eq:respfit}
\end{equation}

\noindent where $a$ and $m$ are free parameters to be determined. The energy dependencies of the response shown in Fig.~\ref{fig:linFtfp}(b) and (d) were approximated with Eq.~\ref{eq:respfit}. The fits to data and {\small FTFP\_BERT} in the studied energy range resulted in $\frac{\chi^2}{\mathrm{NDF}} < 1$. The response predicted by the {\small QGSP\_BERT} physics list exhibits variations in the model transition region around 10--15~GeV and the fits give $\frac{\chi^2}{\mathrm{NDF}} > 5$. The values of the parameters $a$ and $m$, obtained from fits to pion and proton responses for data and {\small FTFP\_BERT}, are shown in Table~\ref{tab:respfit}. 

The power-law parametrisation provides enough flexibility to describe a wide range of dependencies with different representations of response, for instance for protons. In the power-law context, the parameter $m$ is expected to be the same for pions and protons~\cite{Groom:2007}. This assumption is supported by the {\small FTFP\_BERT} physics list, for which the value of $m$ for pions agrees within uncertainties with the value of $m$ for protons, extracted from the dependence on available energy. The value of the parameter $m$ for pions from data is higher than the estimate from {\small FTFP\_BERT}. Both parameters for protons, extracted from data and {\small FTFP\_BERT}, agree within uncertainties. 

\begin{table}
 \caption{Parameters $m$ and $a$, obtained from the fit of Eq.~\protect\ref{eq:respfit} to the dependencies of the pion and proton response on available energy for CALICE test beam data and simulation with the {\small FTFP\_BERT} physics list.}
 \label{tab:respfit}
 \begin{center}
  \begin{tabular}{|c|c|c|c|c|}
   \hline
   & \multicolumn{2}{|c|}{$\pi^{+}$}  & \multicolumn{2}{|c|}{proton} \\ 
   \cline{2-5}
    & $m$ & $a$ & $m$ & $a$  \\
   \hline
    Data       & 0.94$\pm$0.04 & 0.19$\pm$0.03  
               & 0.88$\pm$0.04 & 0.28$\pm$0.05 \\
    FTFP\_BERT & 0.83$\pm$0.03 & 0.25$\pm$0.02  
               & 0.86$\pm$0.02 & 0.29$\pm$0.02 \\ 
   \hline
  \end{tabular}
 \end{center}
\end{table}

\begin{figure}
 \centering
 \includegraphics[width=7.5cm]{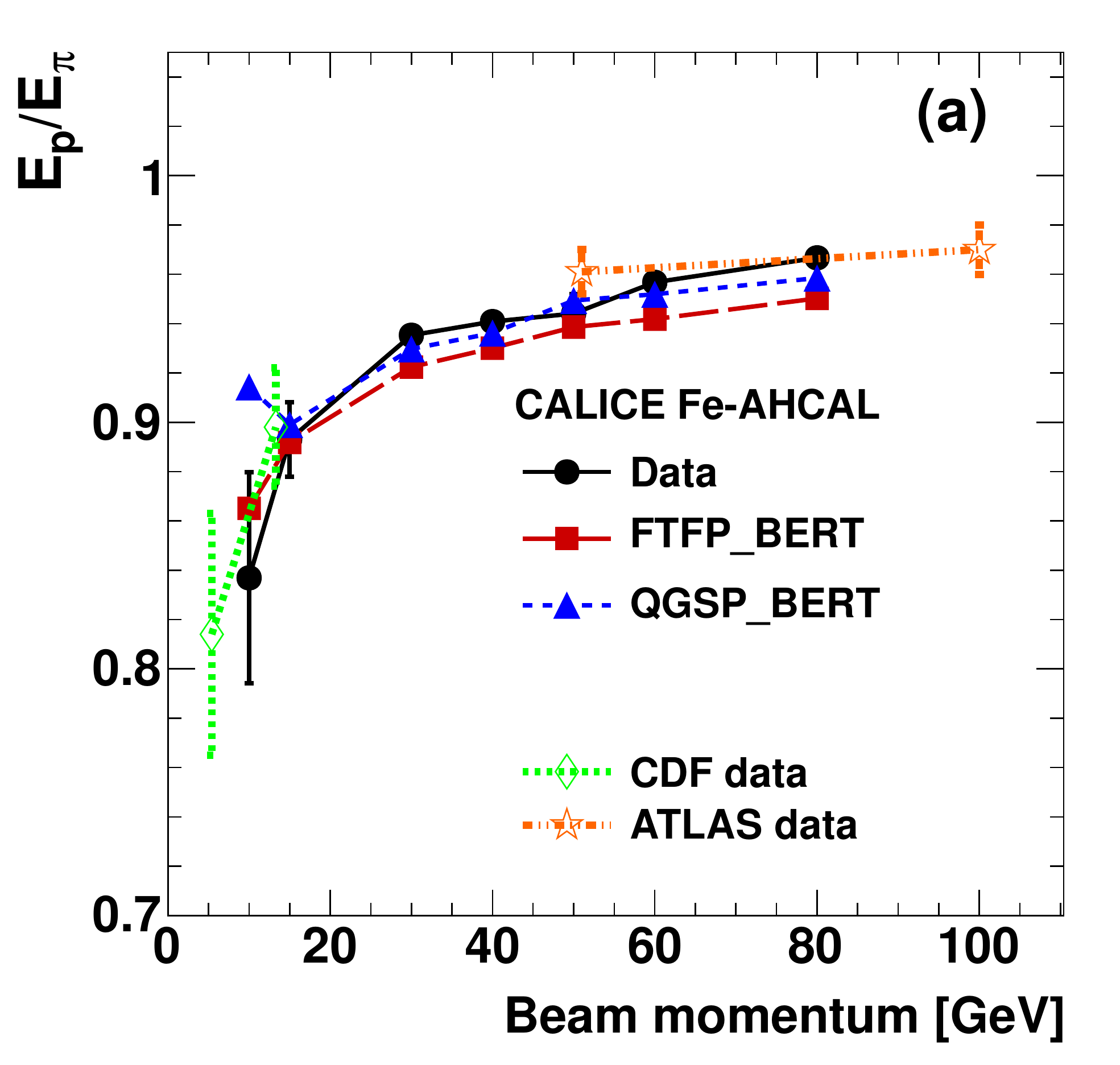}
 \includegraphics[width=7.5cm]{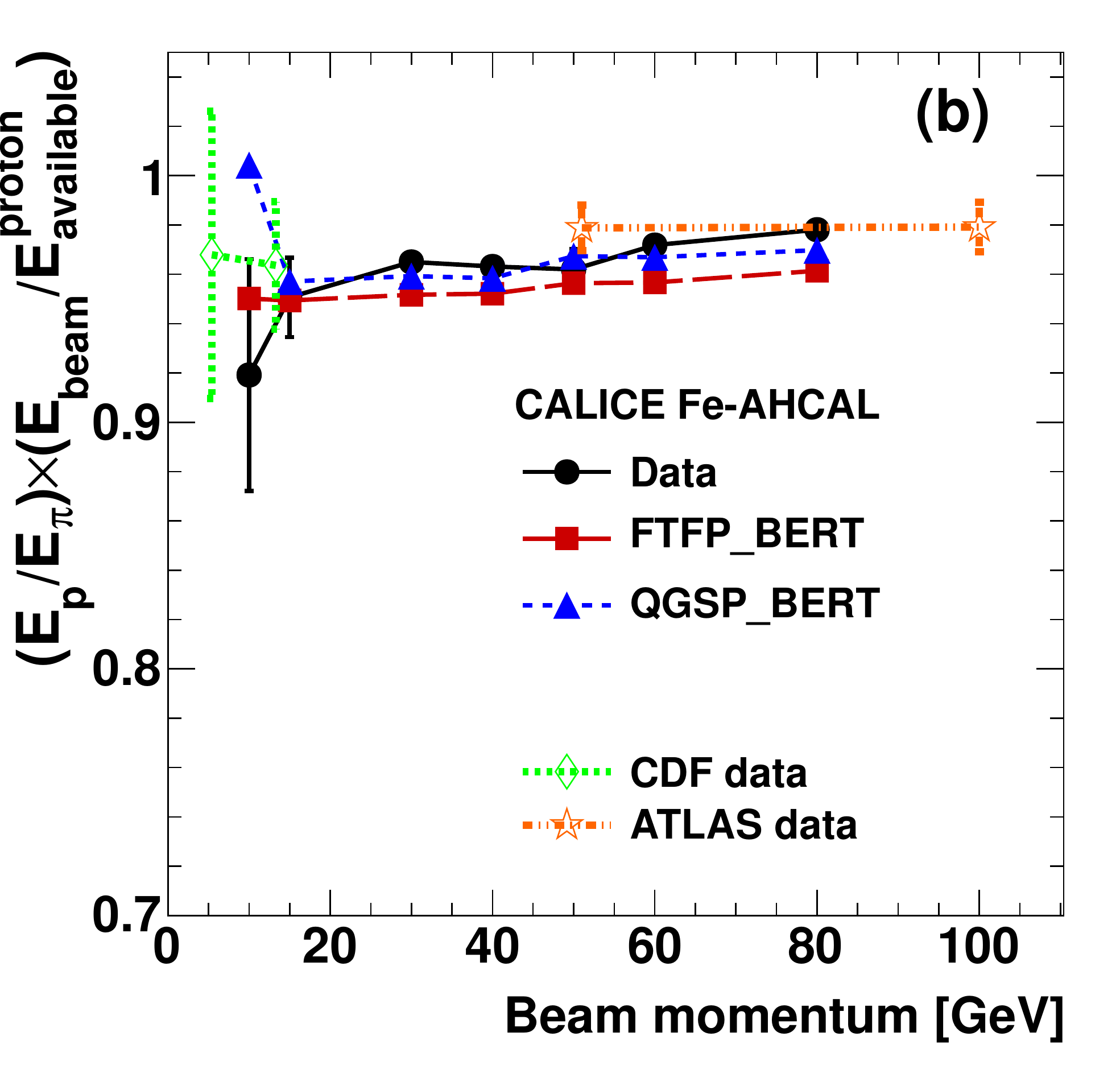}
 \caption{$\mathrm{p}/\pi$ ratio (a) without and (b) with correction for the available energy effect versus beam momentum for data and simulations of the CALICE Fe-AHCAL; error bars show the statistical uncertainties, the mean reconstructed energies are corrected for contamination bias as described in Section~\protect\ref{sec:sys}. The data obtained with the CDF~\cite{CDF:1997} and ATLAS~\cite{ATLAS:2010} hadron calorimeters are shown with open diamonds and stars, respectively.
}
 \label{fig:p2pi}
\end{figure}

The ratio of the calorimeter response of protons to pions of the same initial energy is called the $\mathrm{p}/\pi$ ratio ($\mathrm{p}/\pi = \frac{E_{\mathrm{p}}}{E_{\pi}}$). Figure~\ref{fig:p2pi}(a) shows the $\mathrm{p}/\pi$ ratio extracted in this study for the Fe-AHCAL together with the results obtained for two other iron-scintillator calorimeters: the CDF End Plug hadron calorimeter~\cite{CDF:1997} and the ATLAS Tile hadron calorimeter~\cite{ATLAS:2010}. The results obtained from the measurements performed with different calorimeters are in good agreement.

The energy dependence of the $\mathrm{p}/\pi$ ratio is mainly driven by the difference in measurable energy for mesons and baryons, which dominates below 20~GeV and gives way to other effects at higher energies. This behaviour is qualitatively supported by the comparison of the left and right plots in Fig.~\ref{fig:linFtfp}  
and is quantitatively estimated in Ref.~\cite{CDF:1997}. The available energy effect can be taken into account by multiplying the ratio of reconstructed energies by the ratio of measurable energies $E_{\mathrm{beam}}/E^{\mathrm{proton}}_{\mathrm{available}}$.
The difference between pion and proton response, which remains after taking into account the available energy effect, amounts to 2--5\% as follows from Fig.~\ref{fig:p2pi}(b). This remaining difference is related to the lower probability of $\pi^{0}$ production in the interaction of a proton with a nucleus~\cite{Groom:2007}.

Both physics lists tend to underestimate the $\mathrm{p}/\pi$ ratio above 20~GeV. The {\small FTFP\_BERT} physics list underestimates the $\mathrm{p}/\pi$ ratio due to an overestimate of the pion response while the proton response is reproduced within uncertainties. The predictions of {\small QGSP\_BERT} are closer to the data because both pion and proton response is overestimated by this physics list above 20~GeV. At the same time, abnormal behaviour is visible around the model transition region in the {\small QGSP\_BERT} physics list. 

\subsection{Energy resolution}
\label{sec:resolution}

\begin{figure}
 \centering
 \includegraphics[width=7.5cm]{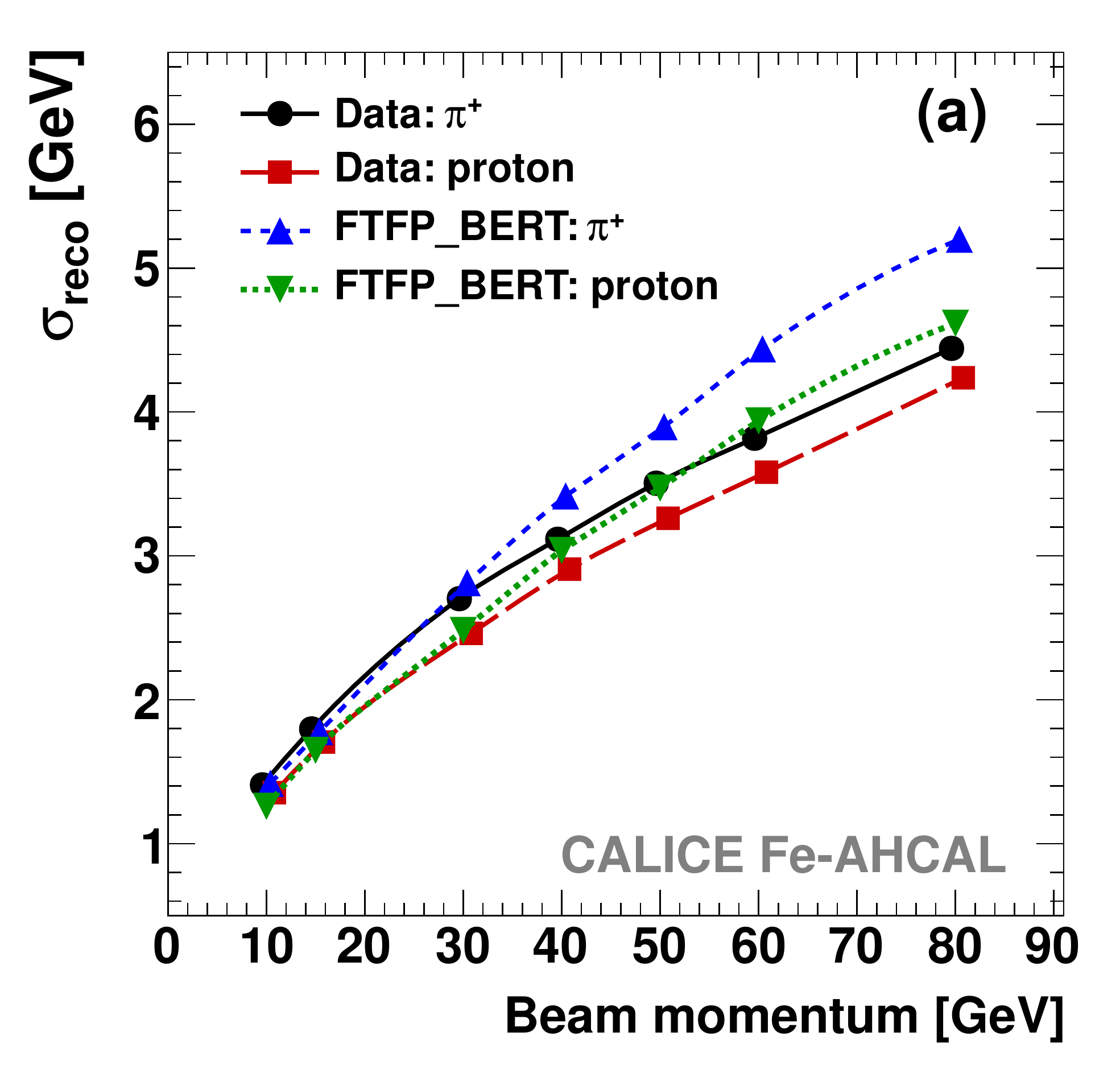}
 \includegraphics[width=7.5cm]{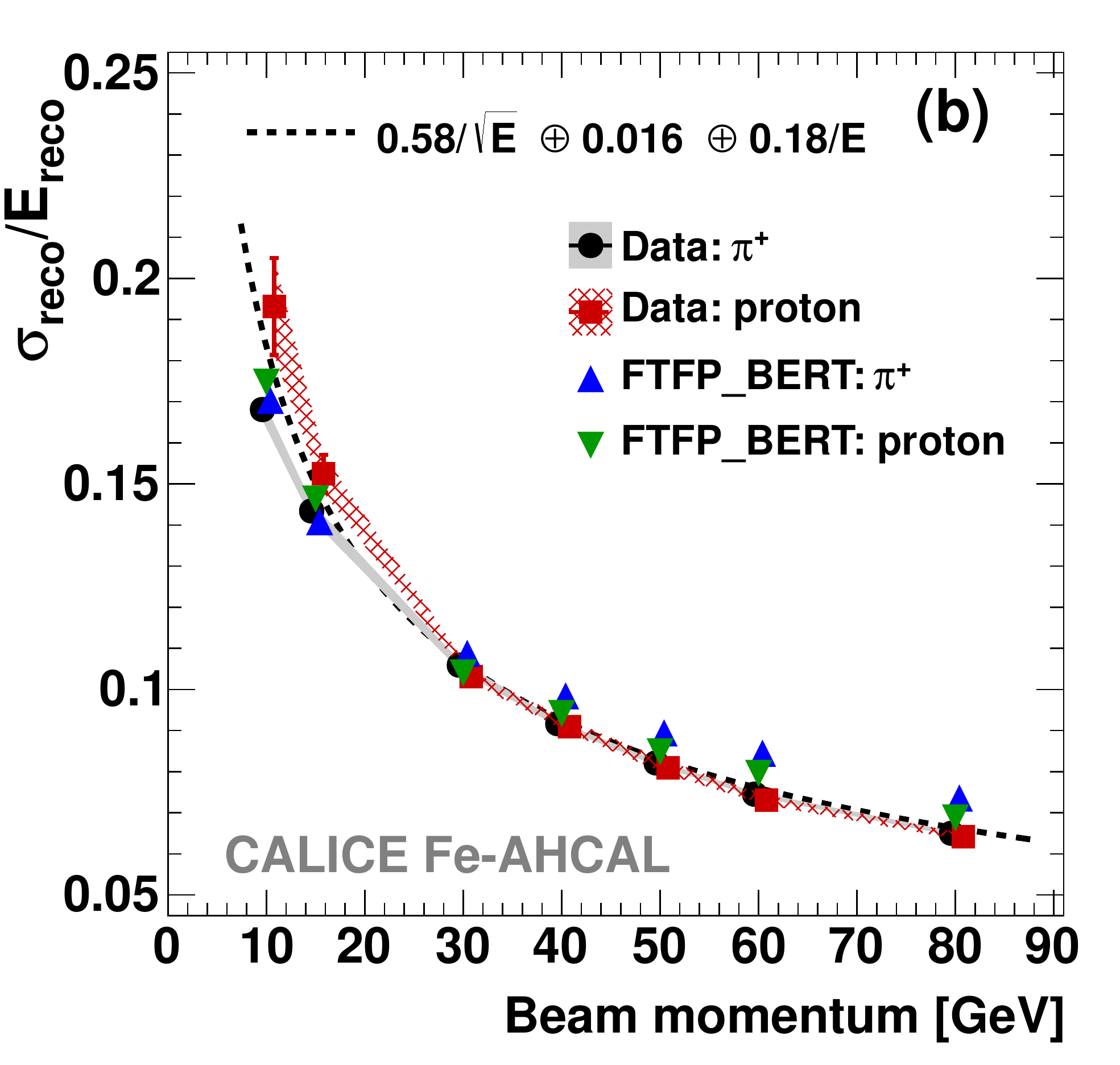}
 \includegraphics[width=7.5cm]{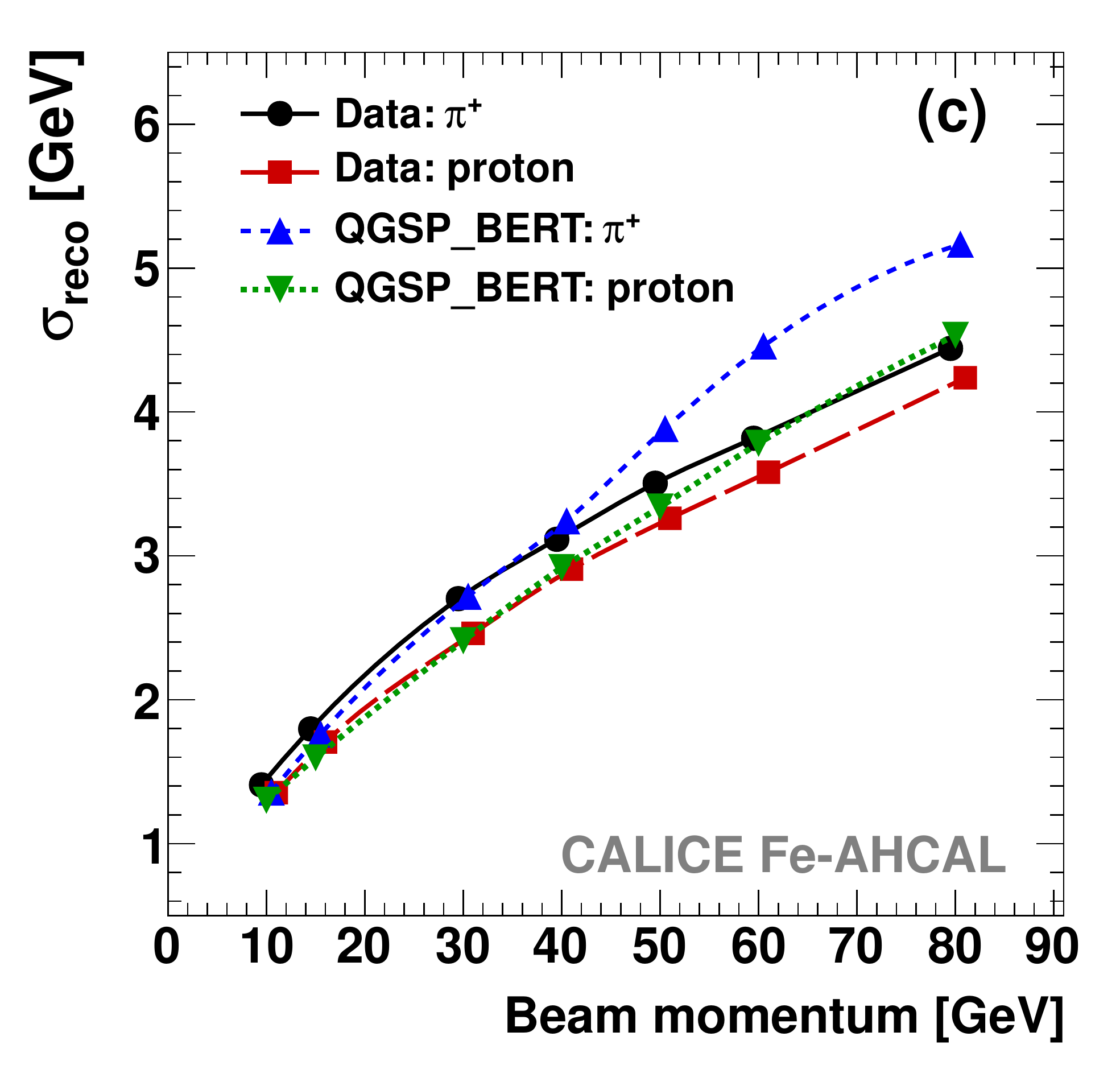}
 \includegraphics[width=7.5cm]{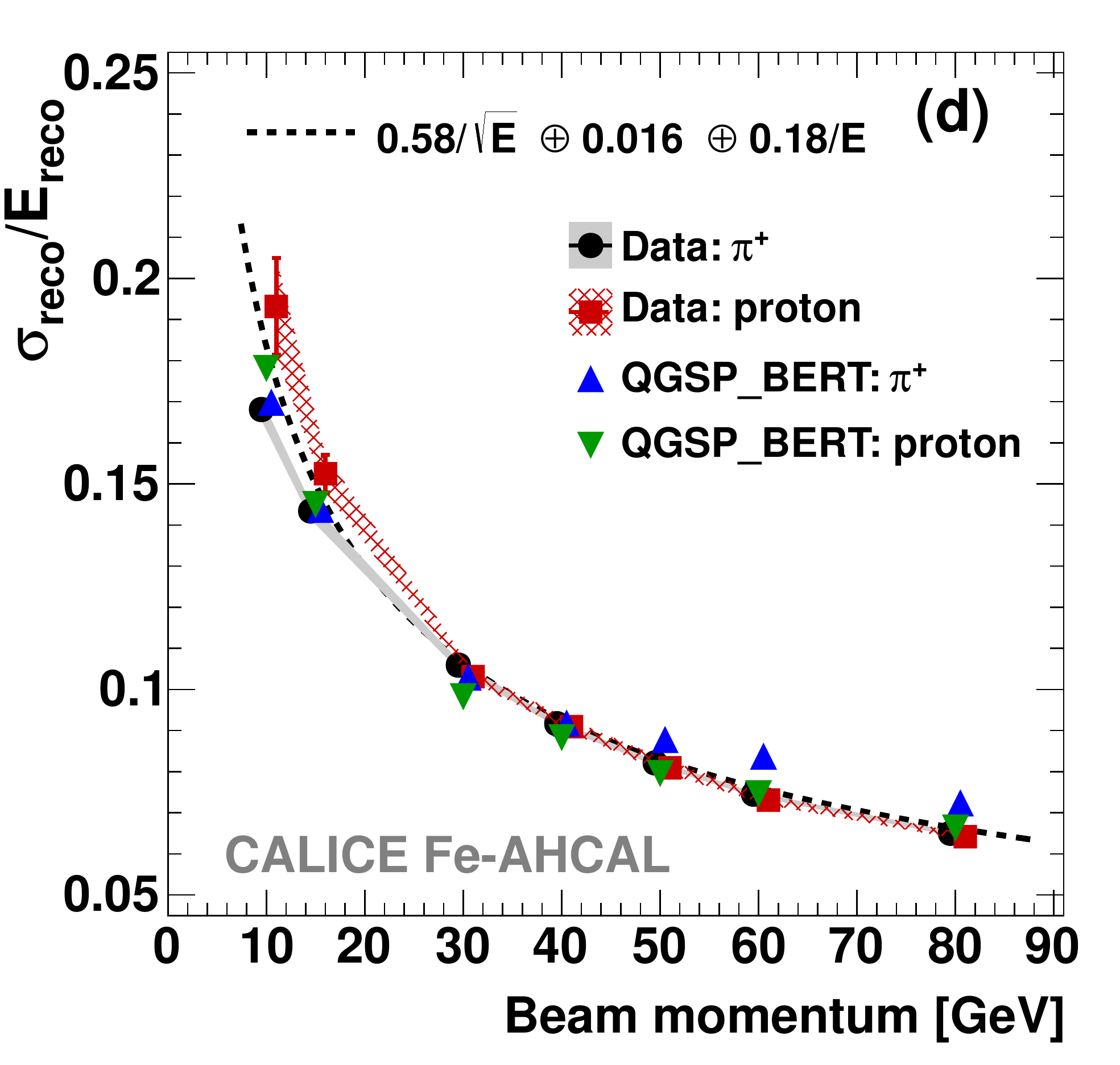}
 \caption{(a,c) Absolute and (b,d) relative energy resolution for pions (circles), protons (squares), simulated pions (triangles), and simulated protons (down triangles) using the {\small FTFP\_BERT} (upper row) and {\small QGSP\_BERT} (bottom row) physics lists. The mean reconstructed energies for data are corrected for contamination bias as described in Section~\protect\ref{sec:sys}. Error bars show the statistical uncertainties. The filled and crosshatched bands show the systematic uncertainties for pion and proton data, respectively (not estimated for the absolute resolution).}
 \label{fig:resFtfp}
\end{figure}

Absolute and relative energy resolutions for pions and protons are shown in Fig.~\ref{fig:resFtfp} for data and simulation with the {\small FTFP\_BERT} and {\small QGSP\_BERT} physics lists. The dashed curves in Fig.~\ref{fig:resFtfp}(b) and (d) represent the result from Ref.~\cite{AHCAL:2012res}, in which the energy dependence of the relative pion energy resolution is parametrised in the energy range 10--80 GeV as a quadratic sum

\begin{equation}
 \frac{\sigma}{E} = \frac{a_1}{\sqrt{E}} \oplus a_2 \oplus \frac{a_3}{E},
 \label{eq:resfunc}
\end{equation}

\noindent where $E$ is in GeV. The stochastic and constant terms, $a_1 = 0.576\pm0.004$~GeV$^{\frac{1}{2}}$ and $a_2 = 0.016\pm0.003$, were obtained in Ref.~\cite{AHCAL:2012res} from the fit to pion data. The noise contribution, $a_3 = 0.18$~GeV, was fixed to the measured noise of the combined calorimeter setup (Si-W~ECAL + Fe-AHCAL + TCMT).

The energy dependence of $\sigma_{\mathrm{reco}}$ for pions is not reproduced by the simulation. The fluctuations of the energy deposition in the simulated hadronic showers grow steeper with increasing energy. This tendency is clearly seen in Fig.~\ref{fig:resFtfp}(a) and (c) where the smooth curves are shown to guide the eyes. The overestimate of $\sigma_{\mathrm{reco}}$ exceeds 15\% at 80~GeV. 
As both response and absolute resolution for pions are overestimated by both physics lists, the behaviour of the relative energy resolution is well reproduced by simulations.

The absolute energy resolution tends to be better for protons than for pions. The simulations predict a larger difference in $\sigma_{\mathrm{reco}}$ between pions and protons than is observed in data. The relative energy resolutions for protons and pions are within uncertainties.

\subsection{Longitudinal shower depth}
\label{sec:lcog}

\begin{figure}
 \centering
 \includegraphics[width=7.5cm]{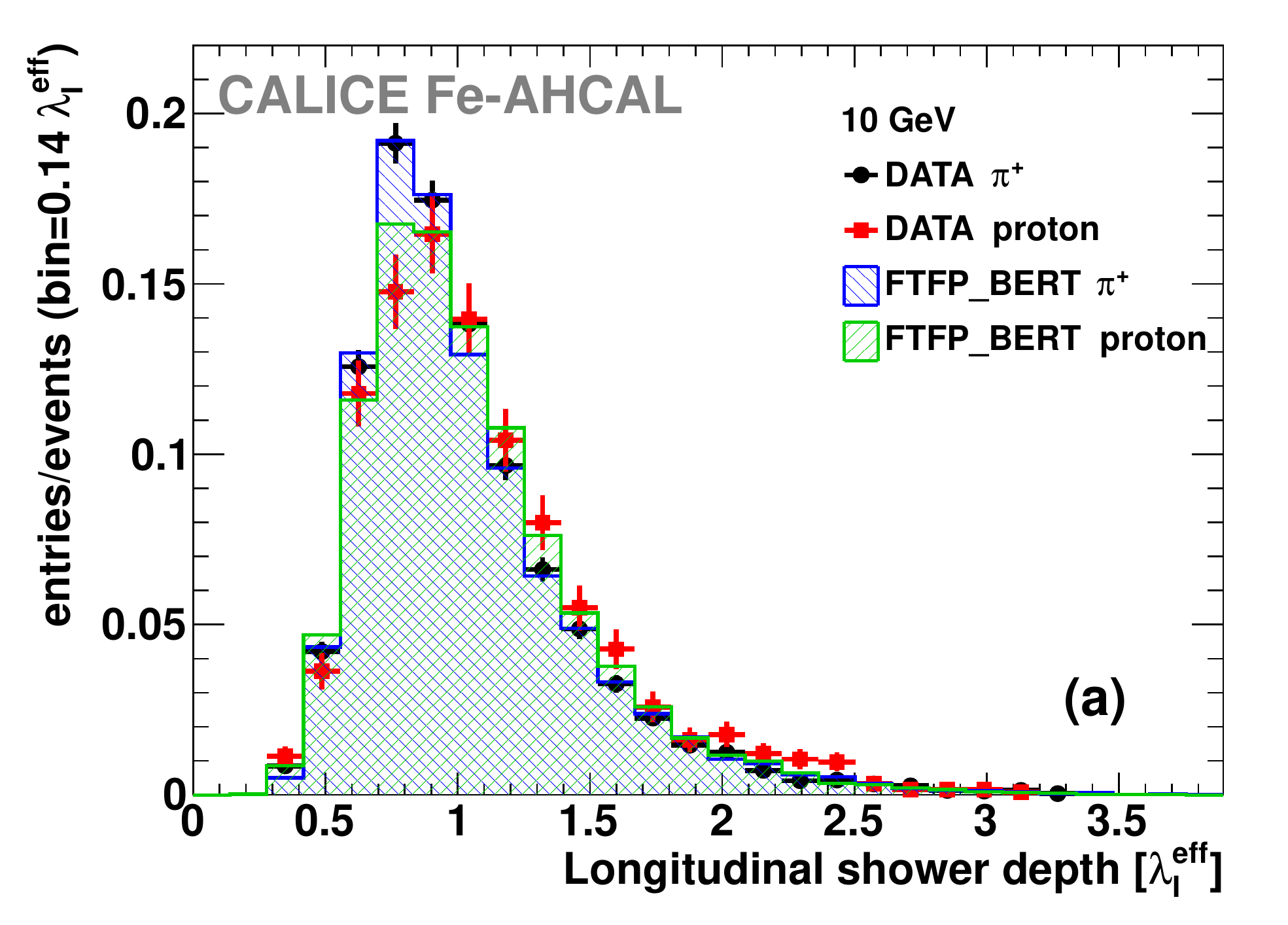}
 \includegraphics[width=7.5cm]{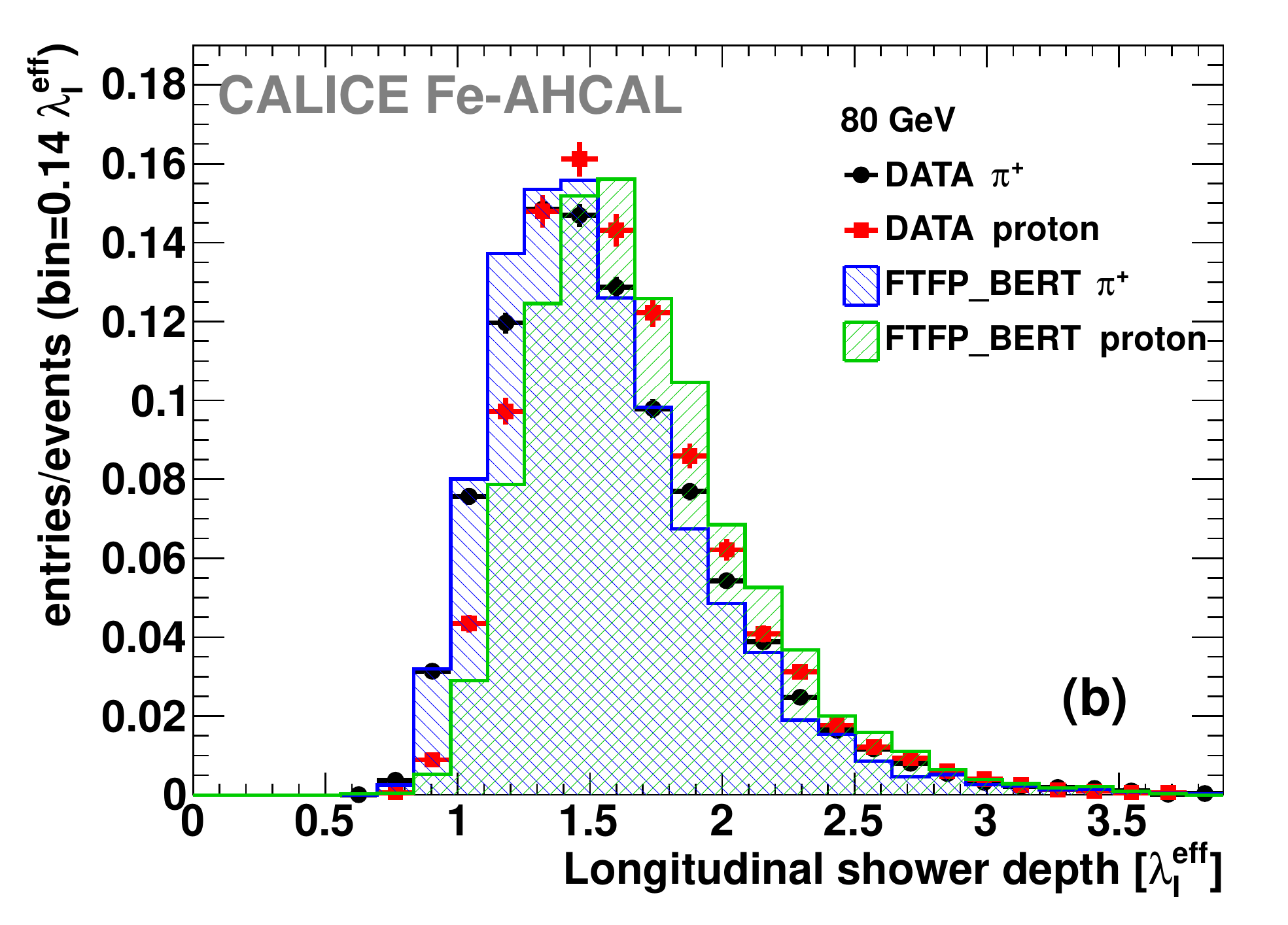}
 \caption{Distributions of the longitudinal shower depth $Z0$ of pion- and proton-induced showers at initial momentum (a) 10~GeV/$c$ and (b) 80~GeV/$c$ for data (points) and simulations by the {\small FTFP\_BERT} physics list (hatched histograms). Error bars show the statistical uncertainties.}
 \label{fig:Z0dist}
\end{figure}

 The observable $Z0$ does not depend on the shower start position and is convenient for the comparison of showers induced by different types of hadrons with different nuclear interaction lengths. 
Typical distributions of $Z0$ are shown in Fig.~\ref{fig:Z0dist} for pions and protons. The longitudinal shower depth of a pion shower tends to be closer to the shower start than that of a proton shower. 

\begin{figure}
 \centering
 \includegraphics[width=7.5cm]{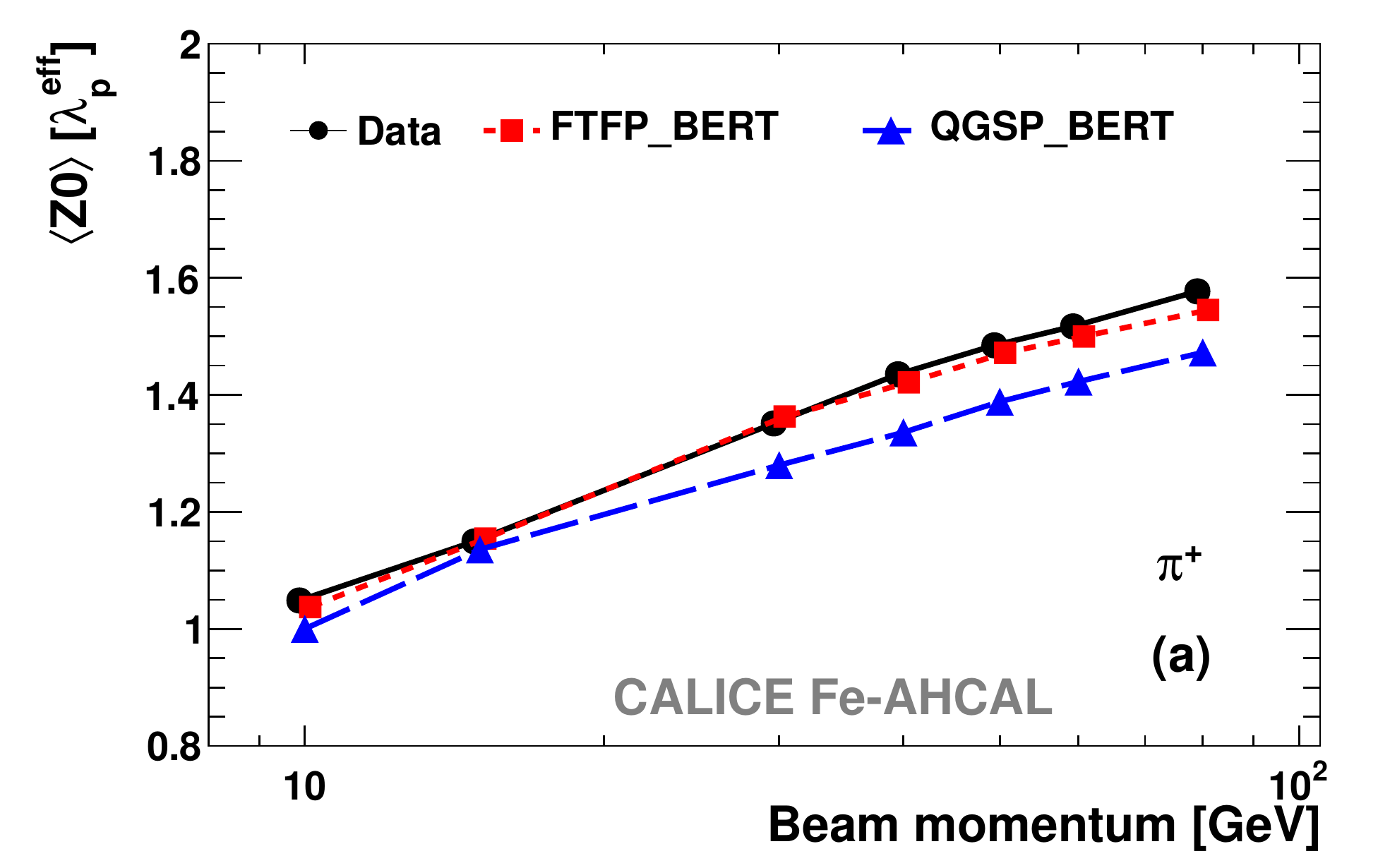}
 \includegraphics[width=7.5cm]{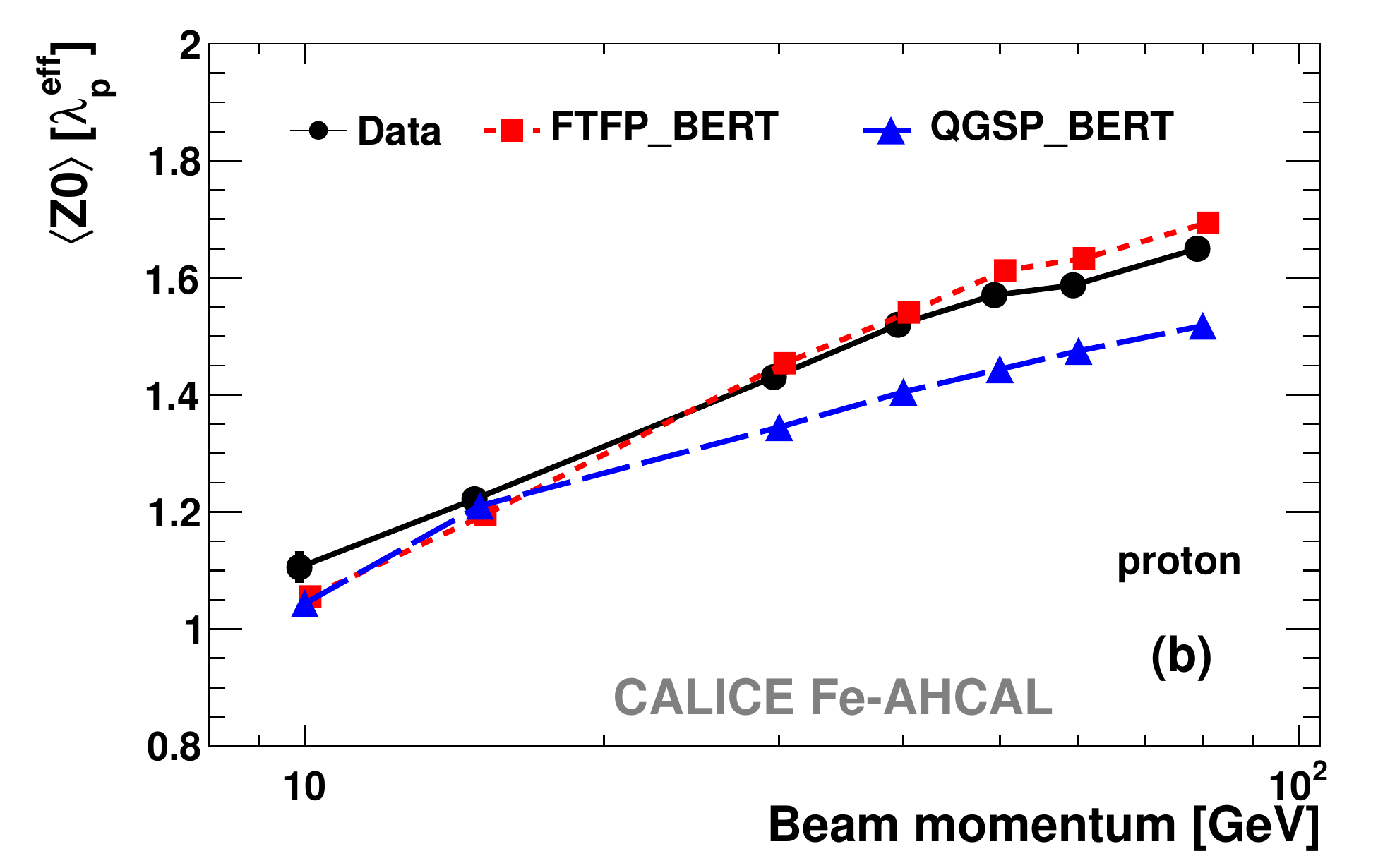}
 \caption{Mean longitudinal shower depth of (a) pion and (b) proton-induced showers in units of $\lambda^{\mathrm{eff}}_{\mathrm{p}}=231$~mm for data (circles, solid lines) and simulations with the {\small FTFP\_BERT} (squares, dotted lines) and {\small QGSP\_BERT} (triangles, dashed lines) physics lists. The values of $\langle Z0 \rangle$ for data are corrected for contamination bias as described in Section~\protect\ref{sec:sys}.
 }
 \label{fig:Z0}
\end{figure}

\begin{figure}
 \centering
 \includegraphics[width=7.5cm]{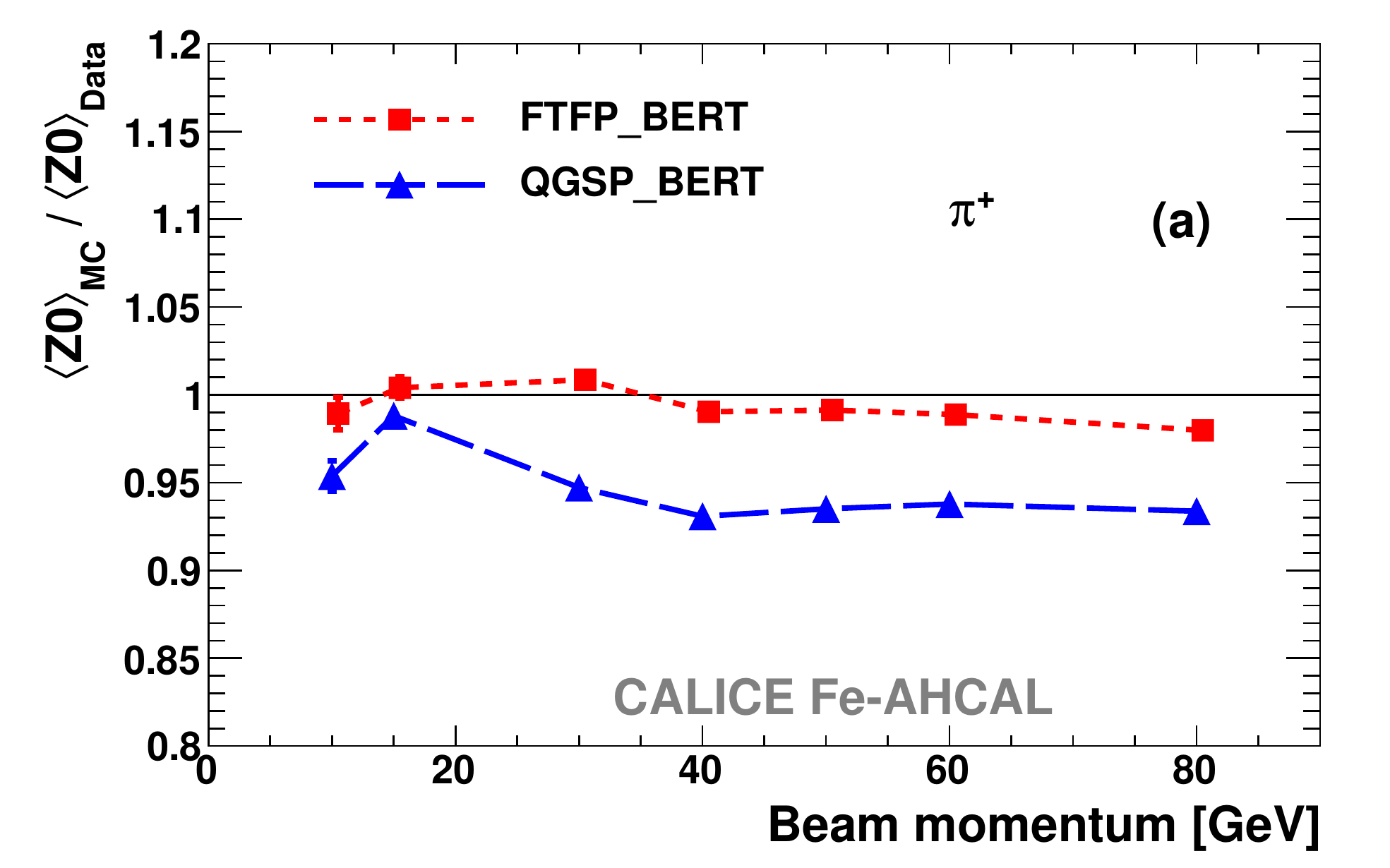}
 \includegraphics[width=7.5cm]{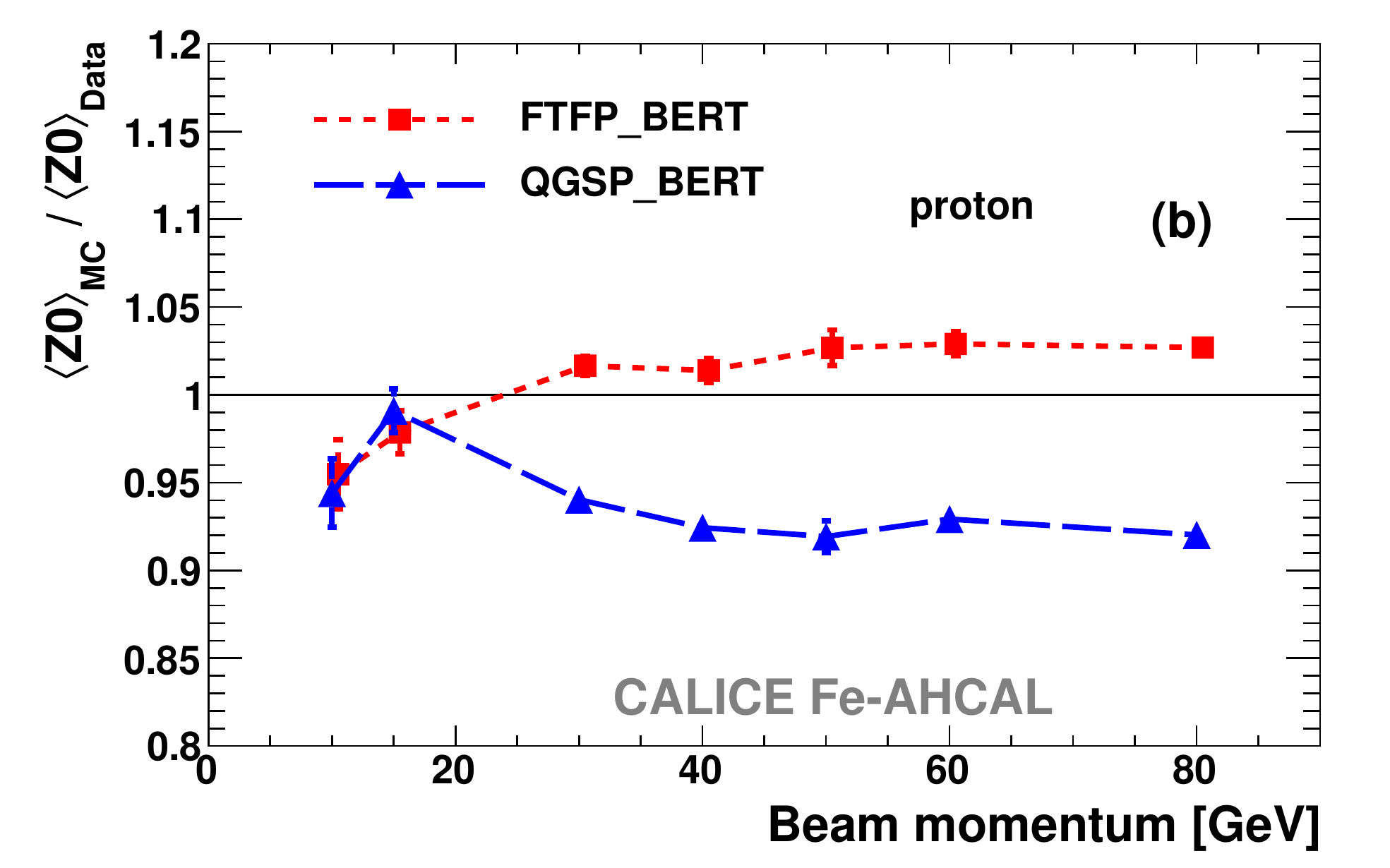}
 \caption{Ratio of the mean longitudinal shower depth for simulations using the {\small FTFP\_BERT} (squares, dotted lines) and {\small QGSP\_BERT} (triangles, dashed lines) physics list to data for (a) pion and (b) proton-induced showers. The values of $\langle Z0 \rangle$ for data are corrected for contamination bias as described in Section~\protect\ref{sec:sys}.
 }
 \label{fig:Z0mc2data}
\end{figure}

The mean longitudinal shower depth $\langle Z0 \rangle$ is extracted from the distributions shown in Fig.~\ref{fig:Z0dist}. The energy dependence of $\langle Z0 \rangle$ in Fig.~\ref{fig:Z0} increases logarithmically with energy from $\sim$1$\lambda^{\mathrm{eff}}_{\mathrm{p}}$ at 10~GeV to $\sim$1.5$\lambda^{\mathrm{eff}}_{\mathrm{p}}$ at 80~GeV.
 Figure \ref{fig:Z0mc2data} shows the simulation to data ratios. 
The {\small QGSP\_BERT} physics list underestimates $\langle Z0 \rangle$ by $\sim$5-7\% for both pions and protons above 20~GeV. The {\small FTFP\_BERT} physics list gives a very good prediction of $\langle Z0 \rangle$ for pions and slightly overestimates the rate of growth for protons.
{\sloppy

}

\begin{figure}
 \centering
 \includegraphics[width=7.5cm]{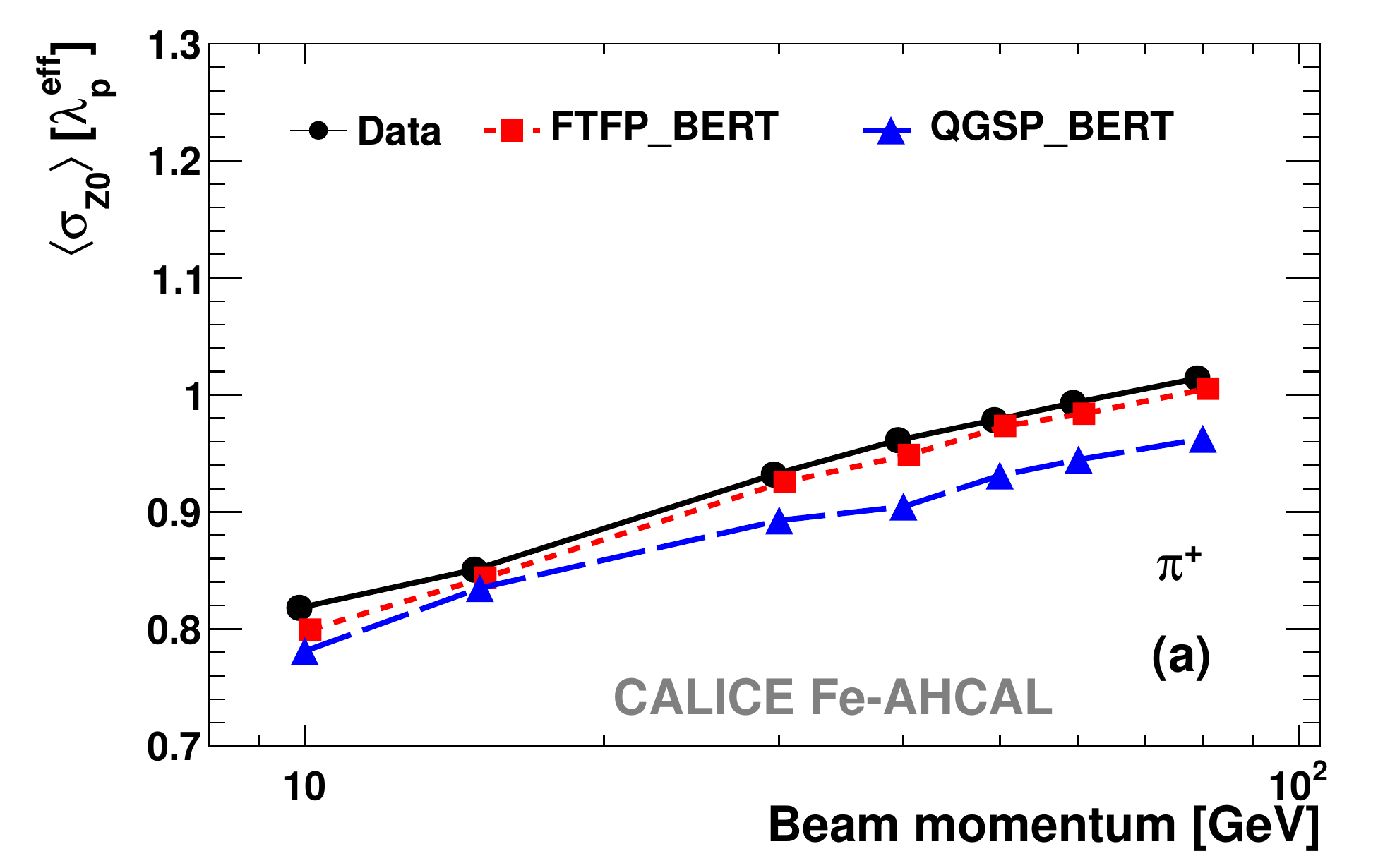}
 \includegraphics[width=7.5cm]{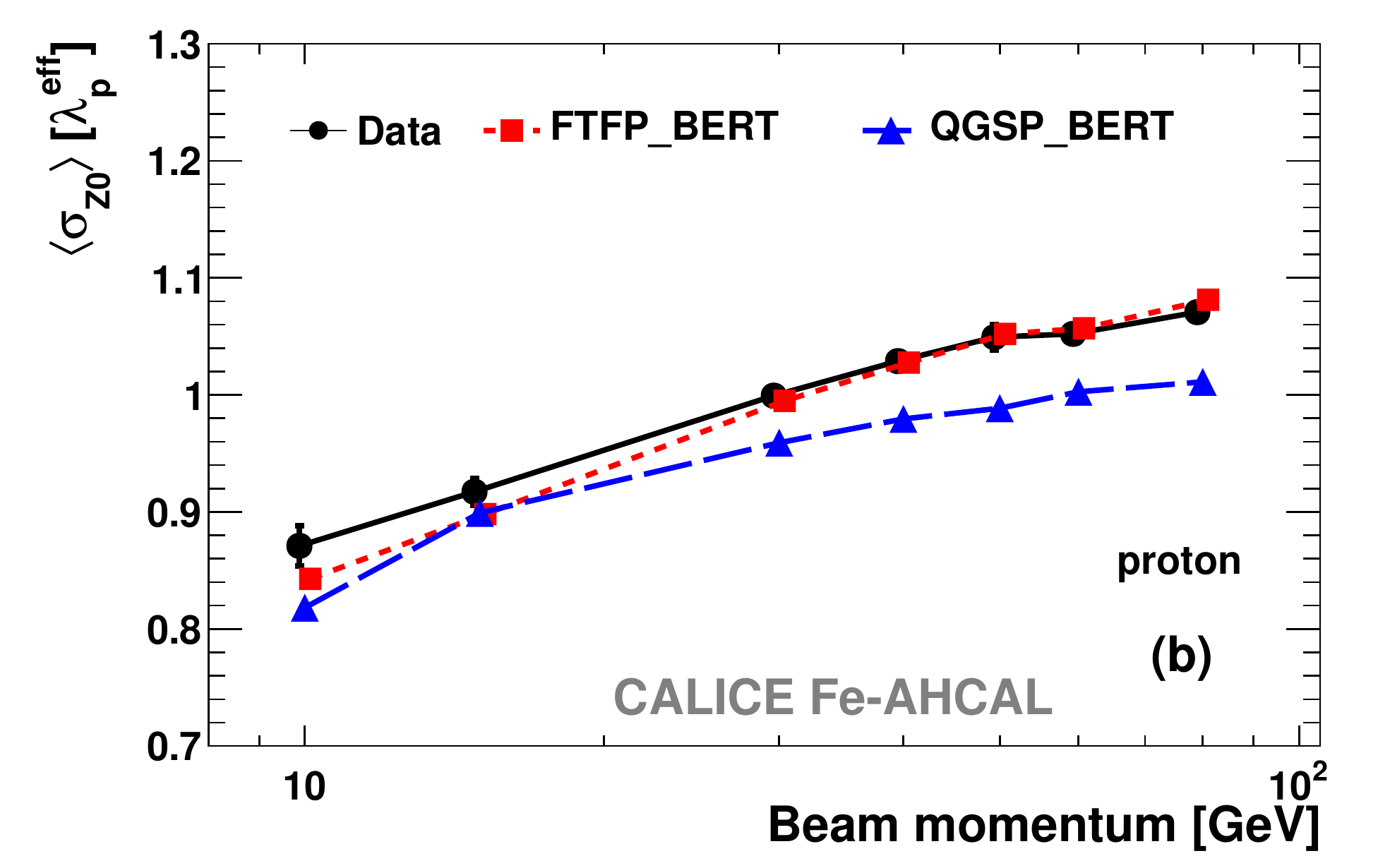}
 \caption{Mean longitudinal dispersion of (a) pion and (b) proton-induced showers in units of $\lambda^{\mathrm{eff}}_{\mathrm{p}}=231$~mm for data (circles, solid lines) and simulations with the {\small FTFP\_BERT} (squares, dotted lines) and {\small QGSP\_BERT} (triangles, dashed lines) physics lists. The values of $\langle \sigma_{Z0} \rangle$ for data are corrected for contamination bias as described in Section~\protect\ref{sec:sys}.
 }
 \label{fig:sigZ0}
\end{figure}

\begin{figure}
 \centering
 \includegraphics[width=7.5cm]{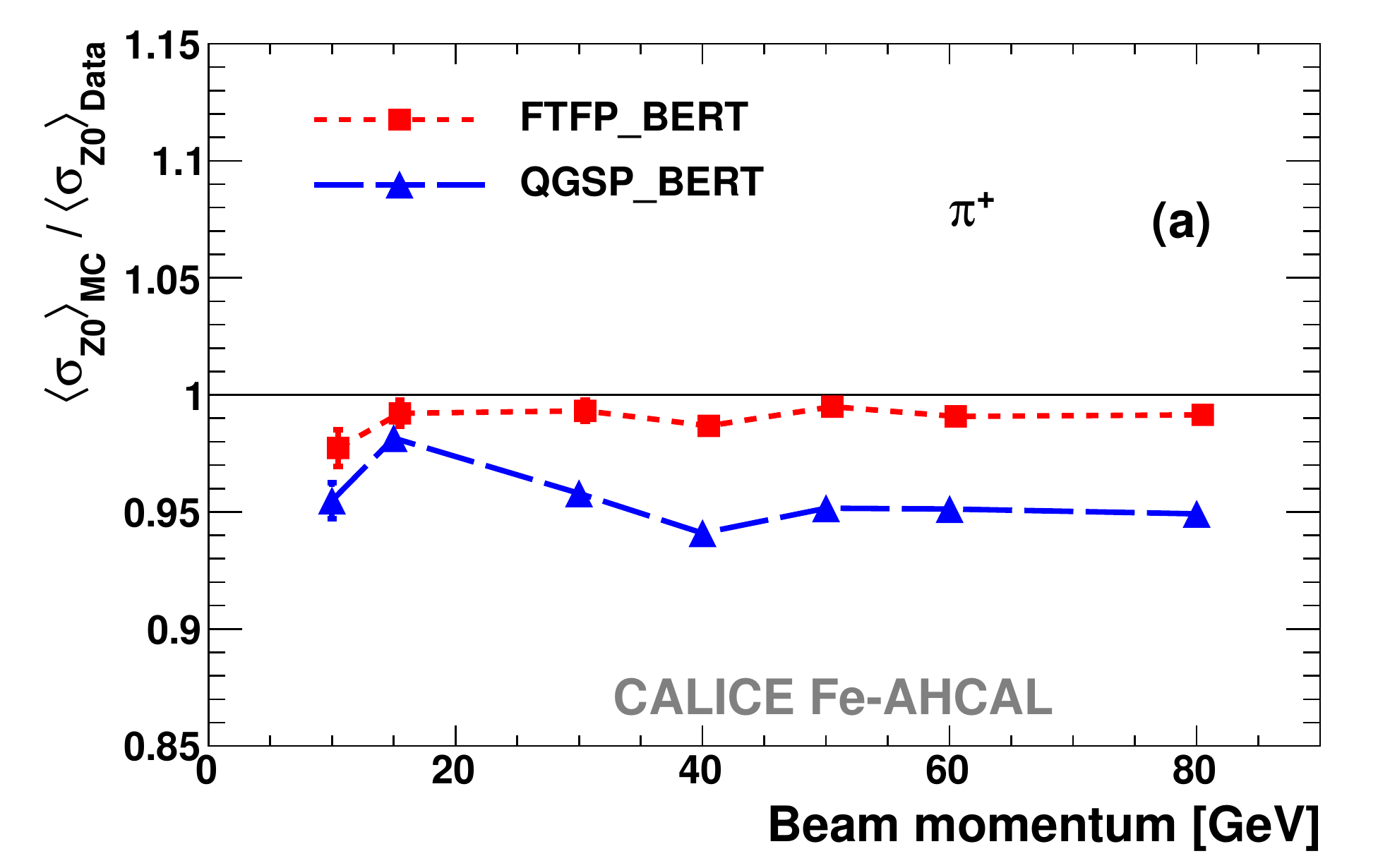}
 \includegraphics[width=7.5cm]{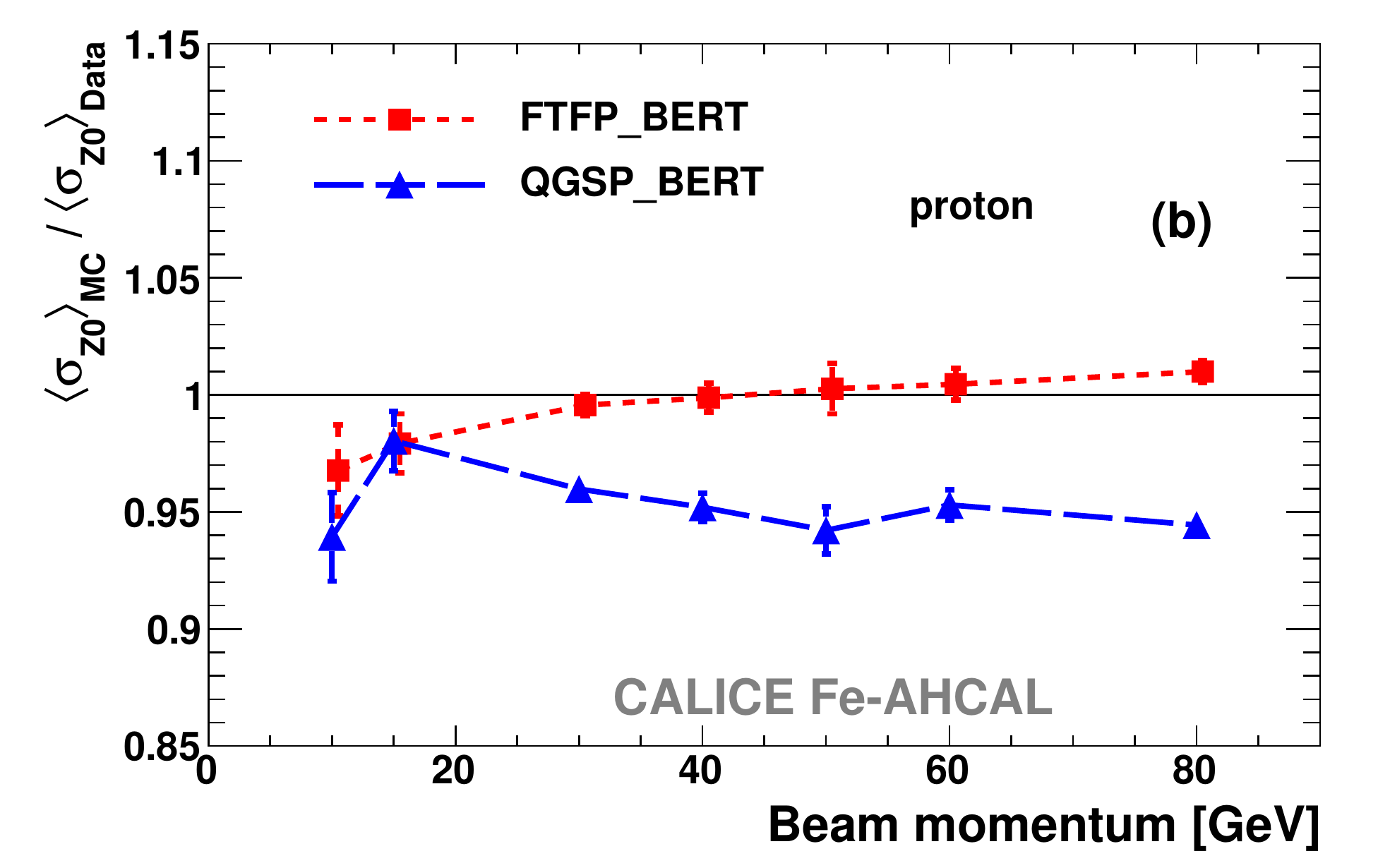}
 \caption{Ratio of the mean longitudinal dispersion for simulations using the {\small FTFP\_BERT} (squares, dotted lines) and {\small QGSP\_BERT} (triangles, dashed lines) physics lists to data for (a) pion and (b) proton-induced showers. The values of $\langle \sigma_{Z0} \rangle$ for data are corrected for contamination bias as described in Section~\protect\ref{sec:sys}.
 }
 \label{fig:sigZ0mc2data}
\end{figure}

The mean longitudinal dispersion $\langle \sigma_{Z0} \rangle$, shown in Fig.~\ref{fig:sigZ0}, is of the same order of magnitude as the mean $\langle Z0 \rangle$ and also increases logarithmically with energy. The values of $\langle \sigma_{Z0} \rangle$ predicted by the {\small FTFP\_BERT} physics list are in agreement with data for both types of hadrons. The {\small QGSP\_BERT} physics list underestimates the mean longitudinal dispersion above 20~GeV by $\sim$5\% (Fig.~\ref{fig:sigZ0mc2data}).

\subsection{Mean shower radius}
\label{sec:mrad}

\begin{figure}
 \centering
 \includegraphics[width=7.5cm]{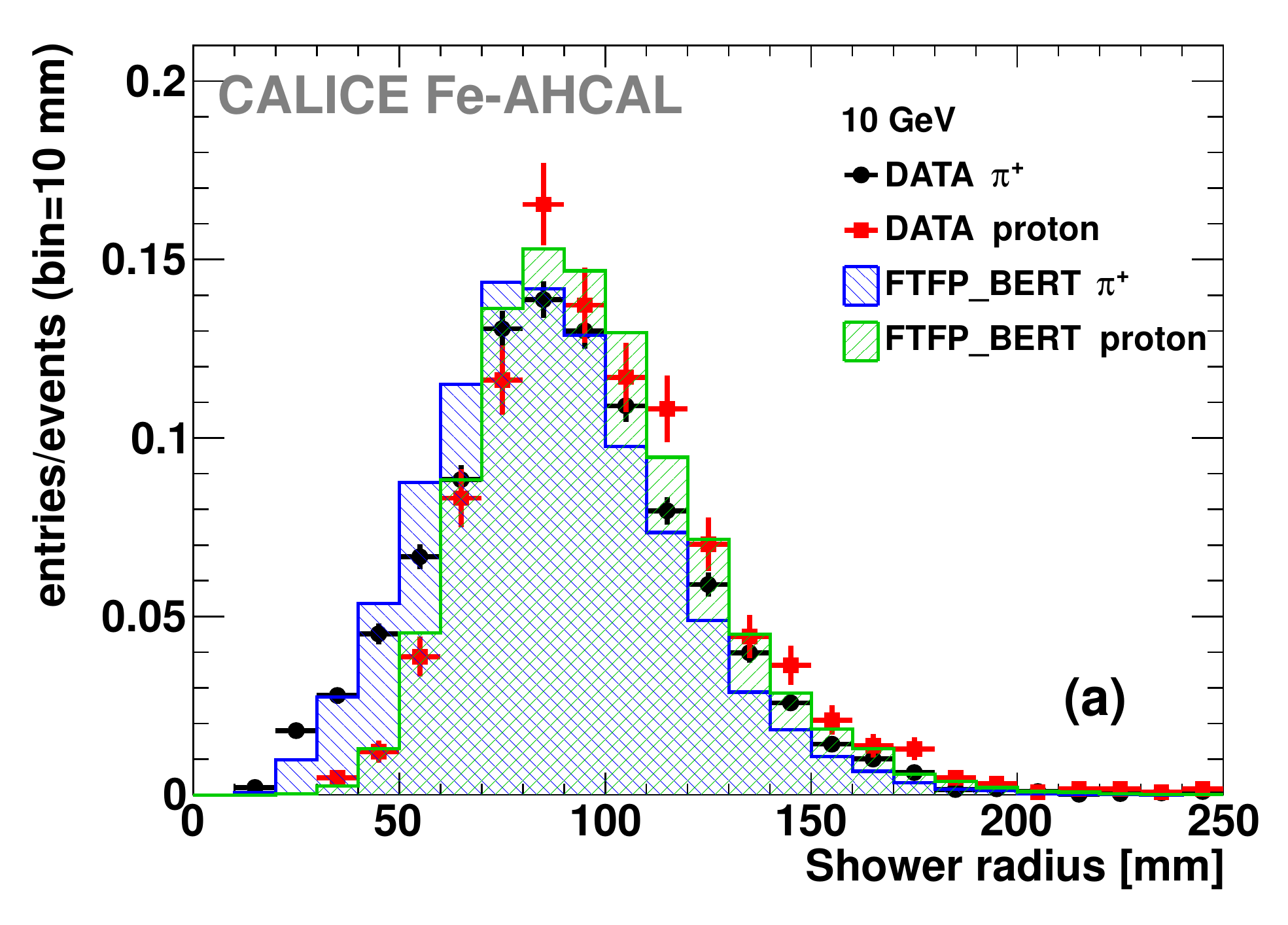}
 \includegraphics[width=7.5cm]{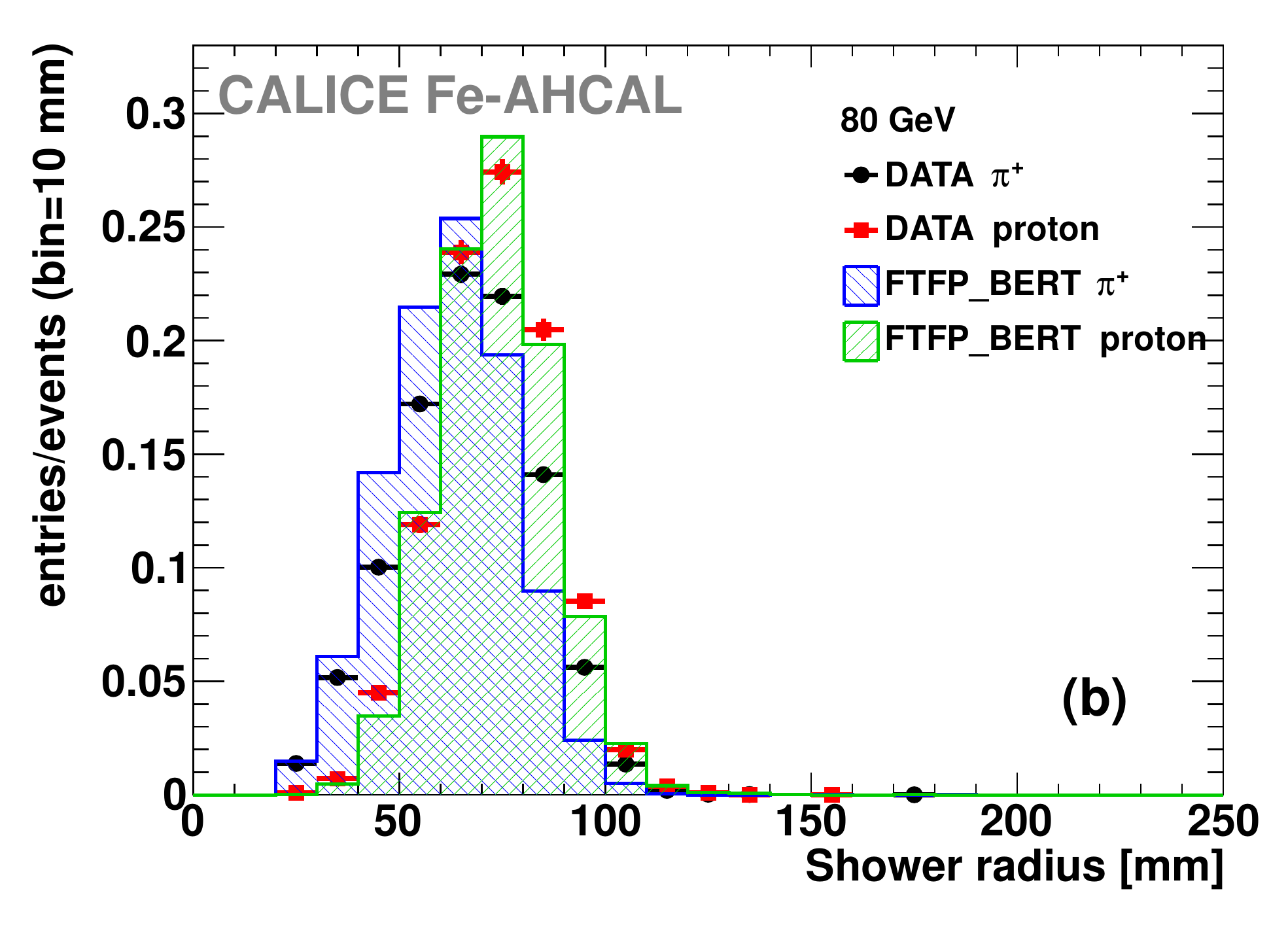}
 \caption{Distributions of the shower radius for pion- and proton-induced showers at initial momentum (a) 10~GeV/$c$ and (b) 80~GeV/$c$ for data (points) and simulation using the {\small FTFP\_BERT} physics list (hatched histograms). Error bars show the statistical uncertainties.}
 \label{fig:Rdist}
\end{figure}

The radial shower development can be characterised by the shower radius $R$ defined in Eq.~\ref{eq:rmean}.
Typical distributions of the shower radius are shown in Fig.~\ref{fig:Rdist}, from which the mean shower radius, $\langle R \rangle$, is extracted for pion- and proton-induced showers. Proton-induced showers tend to be wider than pion showers. For data, the fluctuations of the radius of pion-induced showers are larger than those of proton-induced showers. 

\begin{figure}
 \centering
 \includegraphics[width=7.5cm]{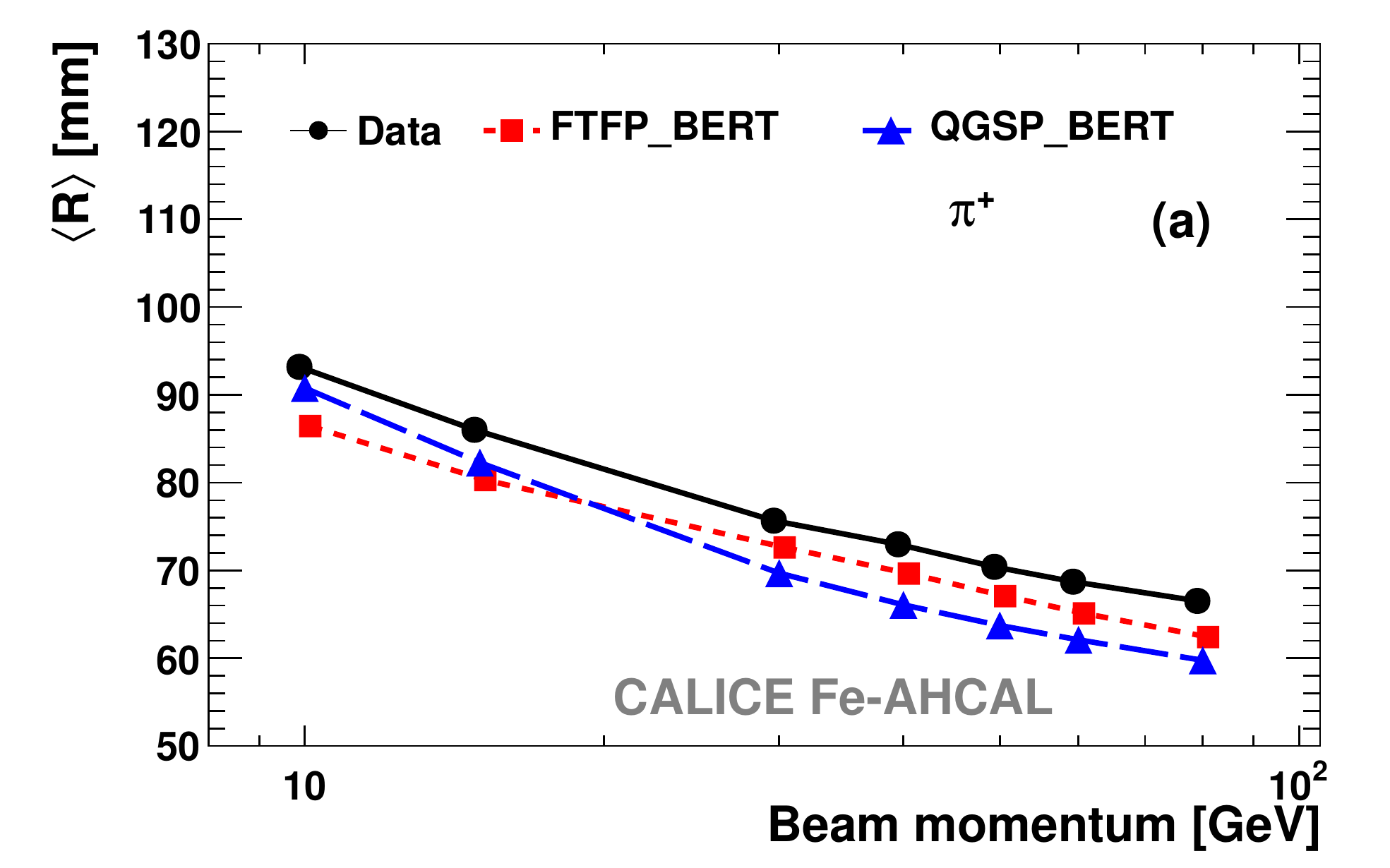}
 \includegraphics[width=7.5cm]{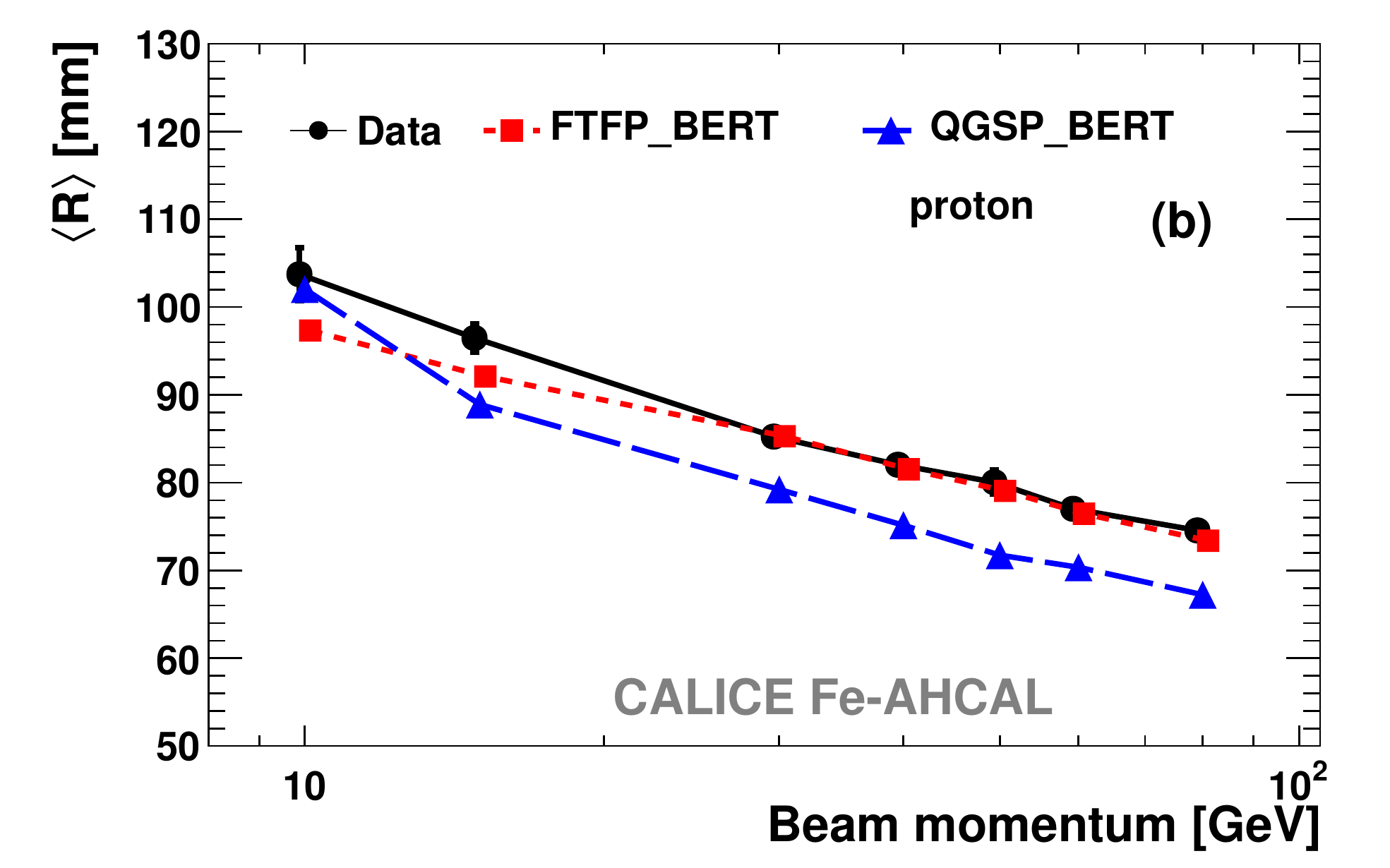}
 \caption{Mean shower radius of (a) pion and (b) proton-induced showers for data (circles, solid lines) and simulations with the {\small FTFP\_BERT} (squares, dotted lines) and {\small QGSP\_BERT} (triangles, dashed lines) physics lists. The values of $\langle R \rangle$ for data are corrected for contamination bias as described in Section~\protect\ref{sec:sys}.
 }
 \label{fig:R0}
\end{figure}

The energy dependencies of the mean shower radius are shown in Fig.~\ref{fig:R0}. The values of $\langle R \rangle$ decrease logarithmically with increasing energy and this general behaviour is well reproduced by all physics lists studied. The pion (proton) showers are observed to be narrower by $\sim$25\% ($\sim$30\%) at 80~GeV than at 10~GeV. 
The trend is explained by the increase of the electromagnetic fraction in hadronic showers with
increased initial energy of an incoming hadron~\cite{Wigmans:2000,Gabriel:1994}. Since more of the shower energy is electromagnetic and since electromagnetic showers are more compact than hadronic showers, the overall shower structure is more compact as beam energy increases.

\begin{figure}
 \centering
 \includegraphics[width=7.5cm]{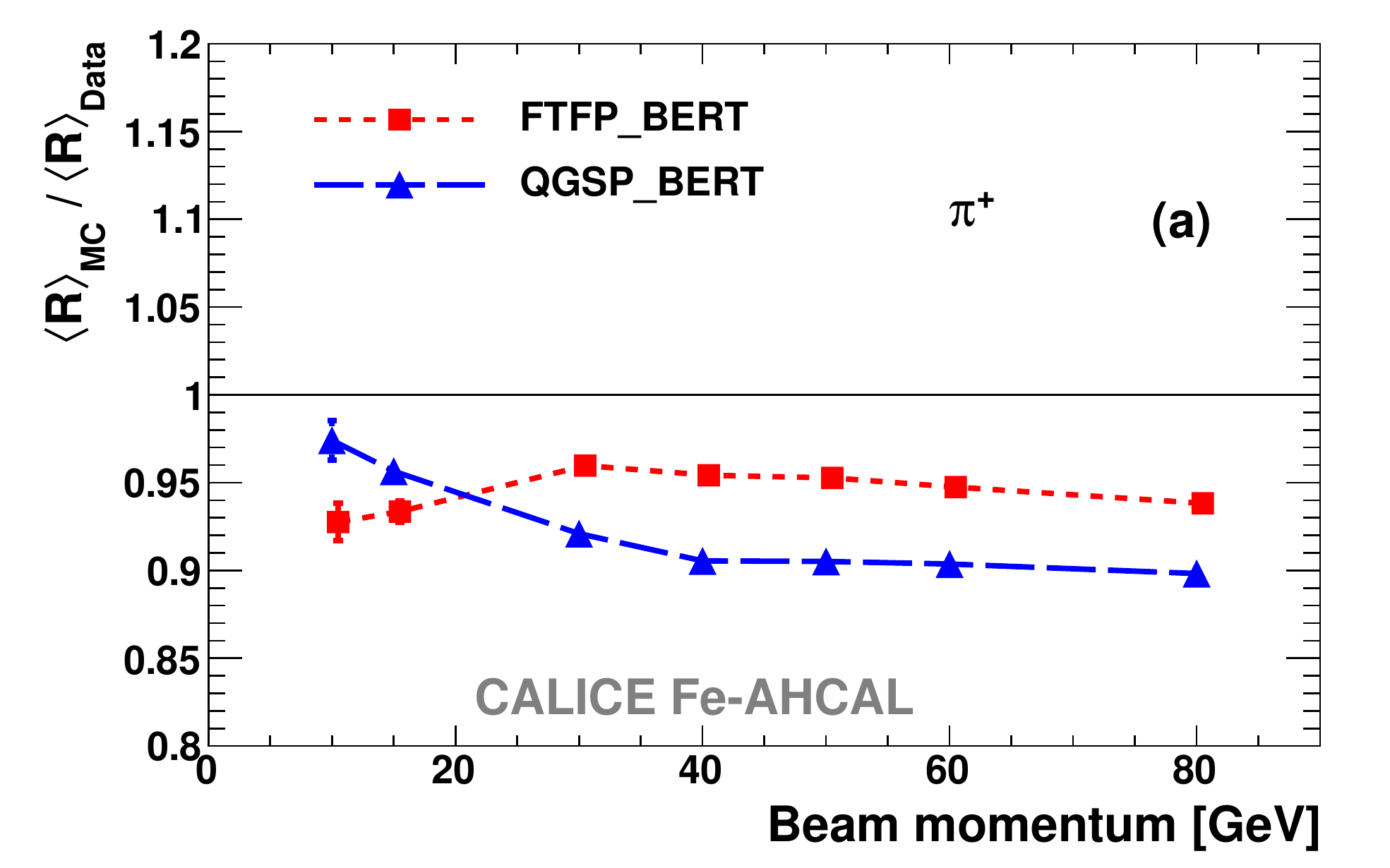}
 \includegraphics[width=7.5cm]{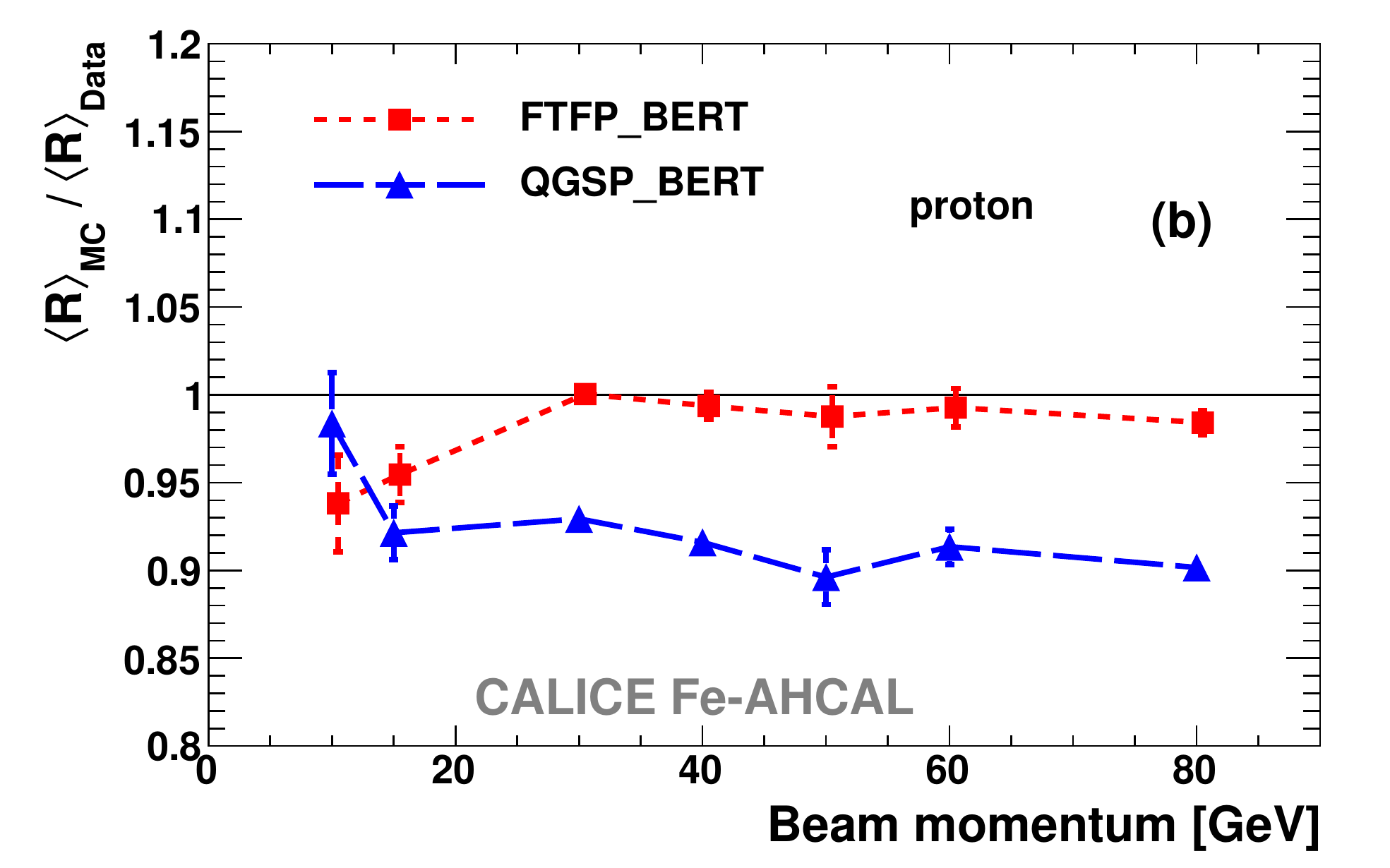}
 \caption{Ratio of the mean shower radius form simulations using the {\small FTFP\_BERT} (squares, dotted lines) and {\small QGSP\_BERT} (triangles, dashes lines) physics lists to data for (a) pion and (b) proton-induced showers. The values of $\langle R \rangle$ for data are corrected for contamination bias as described in Section~\protect\ref{sec:sys}.
 }
 \label{fig:R0mc2data}
\end{figure}

The ratio of simulations to data is shown in Fig.~\ref{fig:R0mc2data}. The {\small FTFP\_BERT} physics list predicts the mean radius of proton showers within uncertainties and underestimates the mean radius of pion showers by $\sim$5-7\%. The {\small QGSP\_BERT} physics list demonstrates better agreement with data at 10~GeV but underestimates the shower width at higher energies by $\sim$10\% for both pions and protons.

\begin{figure}
 \centering
 \includegraphics[width=7.5cm]{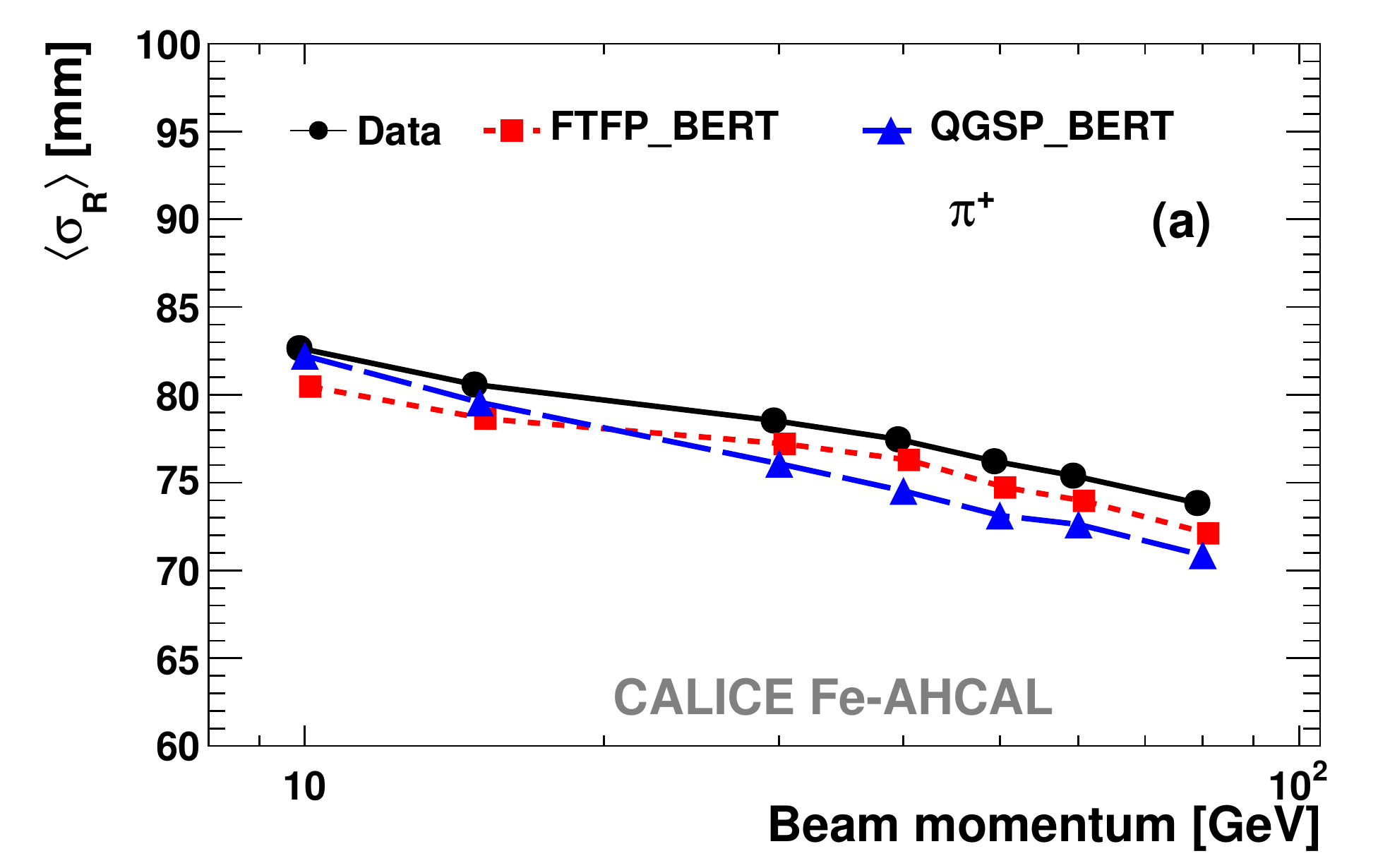}
 \includegraphics[width=7.5cm]{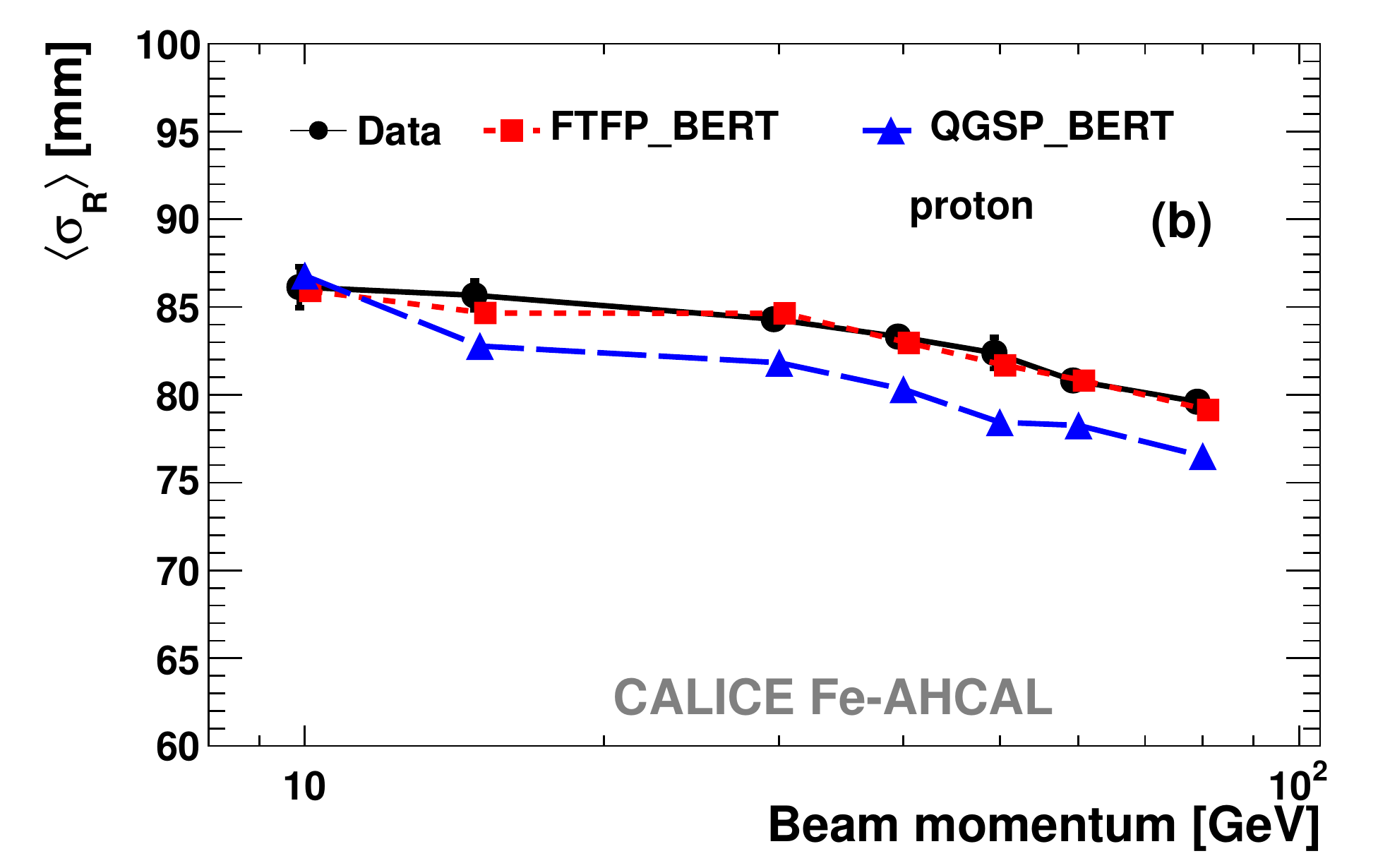}
 \caption{Mean radial dispersion of (a) pion and (b) proton-induced showers for data (circles, solid lines) and simulations with the {\small FTFP\_BERT} (squares, dotted lines) and {\small QGSP\_BERT} (triangles, dashed lines) physics lists. The values of $\langle \sigma_{R} \rangle$ for data are corrected for contamination bias as described in Section~\protect\ref{sec:sys}.
 }
 \label{fig:sigR0}
\end{figure}

\begin{figure}
 \centering
 \includegraphics[width=7.5cm]{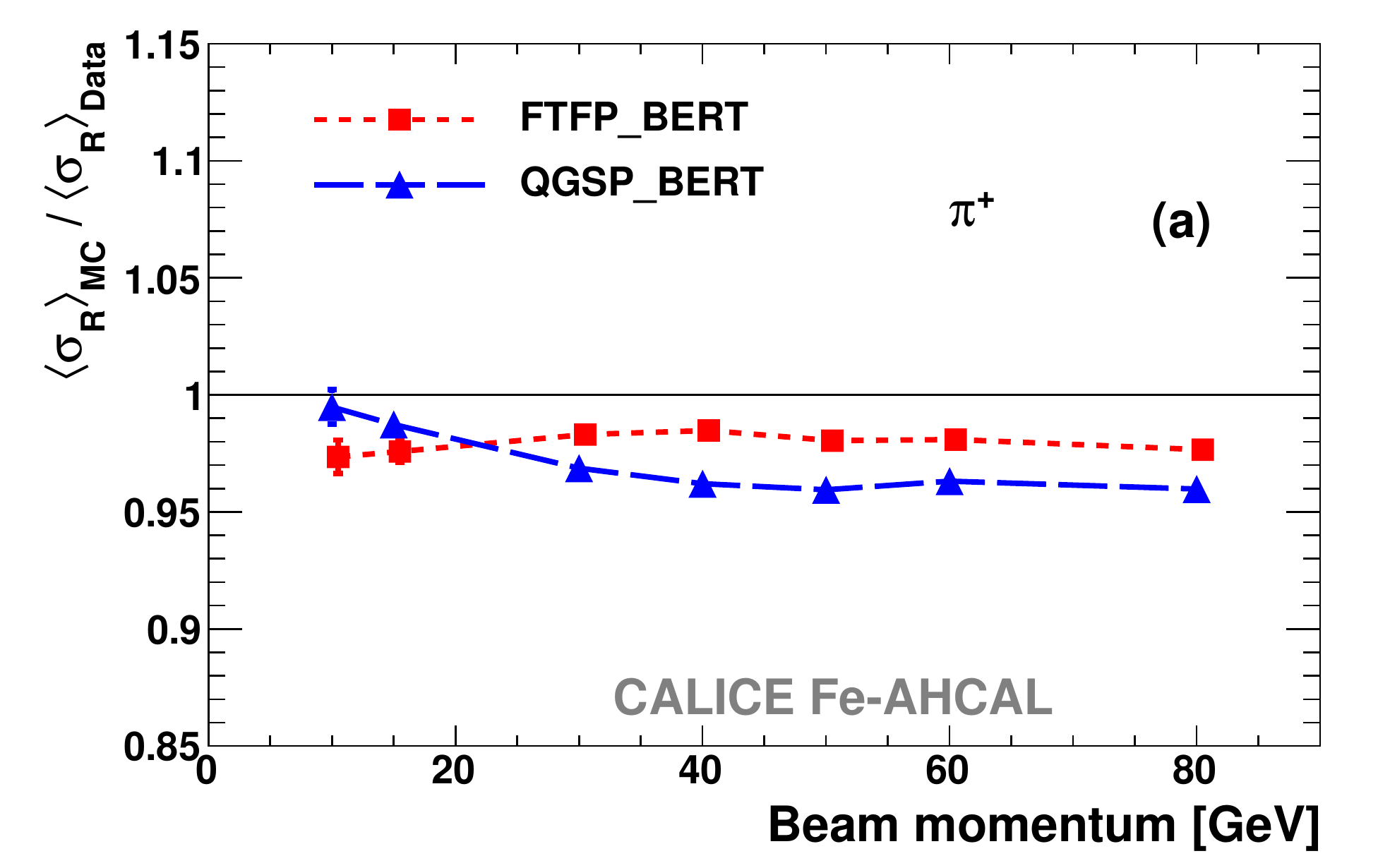}
 \includegraphics[width=7.5cm]{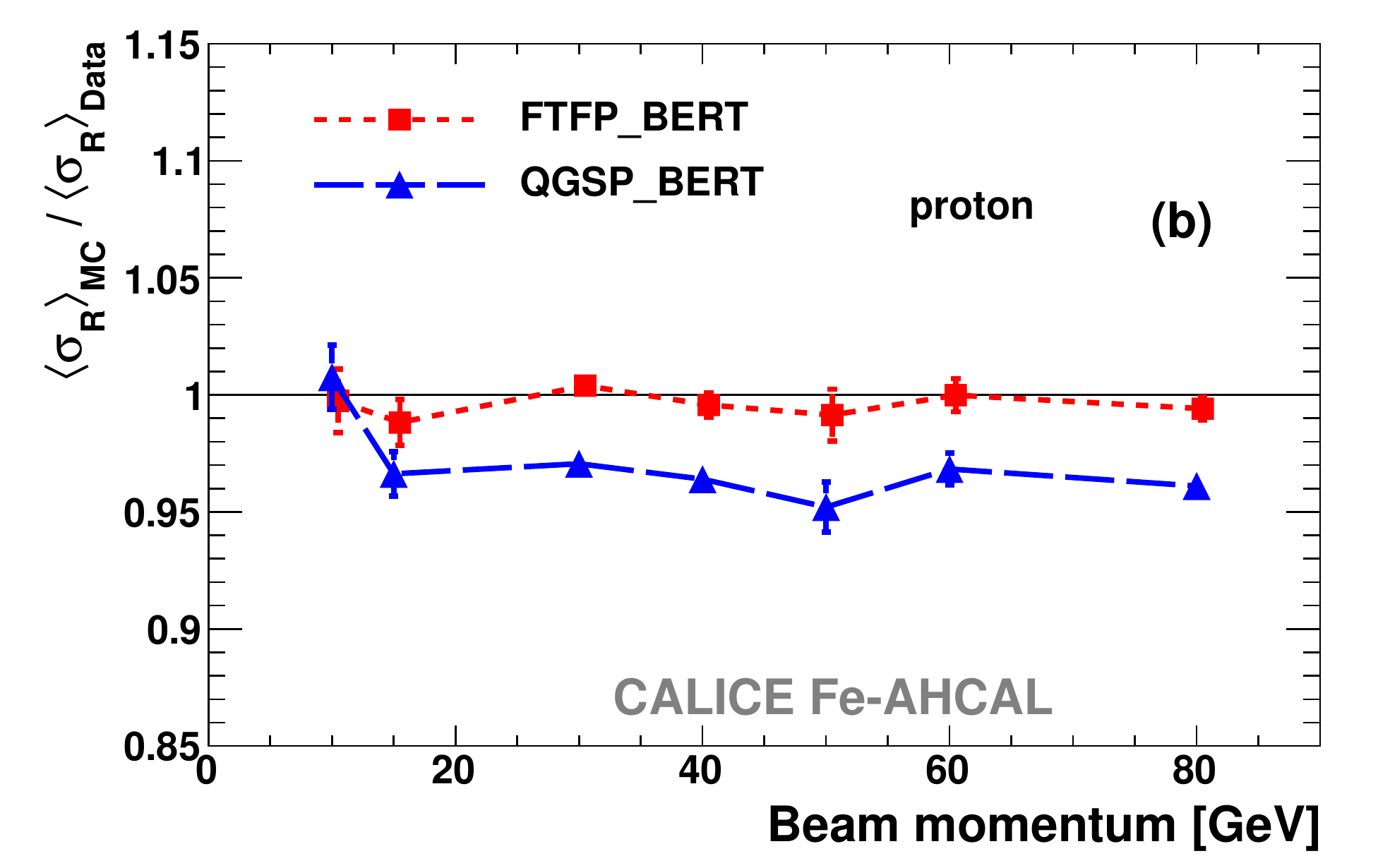}
 \caption{Ratio of the mean radial dispersion from simulations using the {\small FTFP\_BERT} (squares, dotted lines) and {\small QGSP\_BERT} (triangles, dashes lines) physics lists to data for (a) pion and (b) proton-induced showers. The values of $\langle \sigma_{R} \rangle$ for data are corrected for contamination bias as described in Section~\protect\ref{sec:sys}.
 }
 \label{fig:sigR0mc2data}
\end{figure}

The mean radial dispersion $\langle \sigma_{R} \rangle$ is of the same order of magnitude as the mean value $\langle R \rangle$ but decreases more slowly with increasing energy as shown in Fig.~\ref{fig:sigR0}. The discrepancy between data and simulation increases with energy but is smaller than for the mean shower radius. Again, the {\small FTFP\_BERT} physics list describes the data better than {\small QGSP\_BERT}, especially for protons (Fig.~\ref{fig:sigR0mc2data}).

\section{Conclusion}
\label{sec:sum}

Global parameters of showers induced by positive hadrons with initial momenta from 10 to 80~GeV/$c$ in the CALICE analogue scintillator-steel hadronic calorimeter have been analysed and compared with simulations using the {\small QGSP\_BERT} and {\small FTFP\_BERT} physics lists from {\sc Geant4} version 9.6 patch 01. 
In general, the detector response to hadrons tends to increase more rapidly with energy in simulations than in data. Of the two physics lists studied, {\small FTFP\_BERT} gives a better prediction of the response for both pions and protons. The  deficiency of the calorimeter response for protons with respect to pions, which cannot be explained by the difference in available energy, is observed to be $\sim$2--5\%. The ratio of the proton to pion response tends to be underestimated by the simulations.
{\sloppy

}

Proton-induced showers tend to be $\sim$5\% longer and $\sim$10\% wider than pion showers at the same energy in terms of the longitudinal centre of gravity and shower radius, respectively. The spatial shower development for both types of hadrons is much better predicted by the {\small FTFP\_BERT} physics list, especially above 20~GeV. The event-by-event fluctuations of the spatial characteristics are also quite well reproduced. The simulated showers are still narrower than those observed in data but the {\small FTFP\_BERT} physics list predicts the mean shower radius with an accuracy of $\sim$5-7\%. 
  
The most significant discrepancy between test beam data and simulations is seen in the absolute energy resolution for pions. The simulated width of the energy distribution of pion showers increases faster with pion initial energy than observed in data. The relative energy resolutions for pions and protons are in good agreement and are well reproduced by simulations with both physics lists studied.

\acknowledgments

We would like to thank the technicians and the engineers who contributed to the design and construction of the prototypes. We also gratefully acknowledge the DESY and CERN managements for their support and hospitality, and their accelerator staff for the reliable and efficient beam operation. The authors would like to thank the RIMST (Zelenograd) group for their help and sensors manufacturing. 
This work was supported by the 
Bundesministerium f\"{u}r Bildung und Forschung, Germany;
by the  the DFG cluster of excellence `Origin and Structure of the Universe' of Germany; 
by the Helmholtz-Nachwuchsgruppen grant VH-NG-206;
by the BMBF, grant no.~05HS6VH1;
by the Alexander von Humboldt Foundation (including Research Award IV, RUS1066839 GSA);
by the Russian Ministry of Education and Science contracts 4465.2014.2 and 14.A12.31.0006 and the Russian Foundation for Basic Research grant 14-02-00873A;
by MICINN and CPAN, Spain;
by CRI(MST) of MOST/KOSEF in Korea;
by the US Department of Energy and the US National Science Foundation;
by the Ministry of Education, Youth and Sports of the Czech Republic under the projects AV0 Z3407391, AV0 Z10100502, LC527  and LA09042  and by the Grant Agency of the Czech Republic under the project 202/05/0653; 
and by the Science and Technology Facilities Council, UK.

\appendix

\section{Shower start identification and estimate of the nuclear interaction length}
\label{app:lambda}

The following algorithm was used to identify the shower start layer, that is the longitudinal position of the primary inelastic interaction of the incoming hadron.
Two values are calculated on a layer-by-layer basis starting from the calorimeter front layer\footnote{The calorimeter front layer is the first layer of the Fe-AHCAL for the samples taken without the electromagnetic calorimeter or the first layer of the Si-W ECAL.}: the moving average $M_{i}$ of visible energy in ten successive layers up to $i$-th layer and the number of hits in the $i$-th layer $N_{i}$. Two conditions are checked: $(M_{i} + M_{i+1}) > M_{\mathrm{thr}}$ and $(N_{i} + N_{i+1}) > N_{\mathrm{thr}}$. 
When both conditions are satisfied the $i$-th layer is considered to be the shower start layer. The thresholds $M_{\mathrm{thr}}$ and $N_{\mathrm{thr}}$ were optimised using the simulated pion samples.

\begin{figure}
 \centering
 \includegraphics[width=7.5cm]{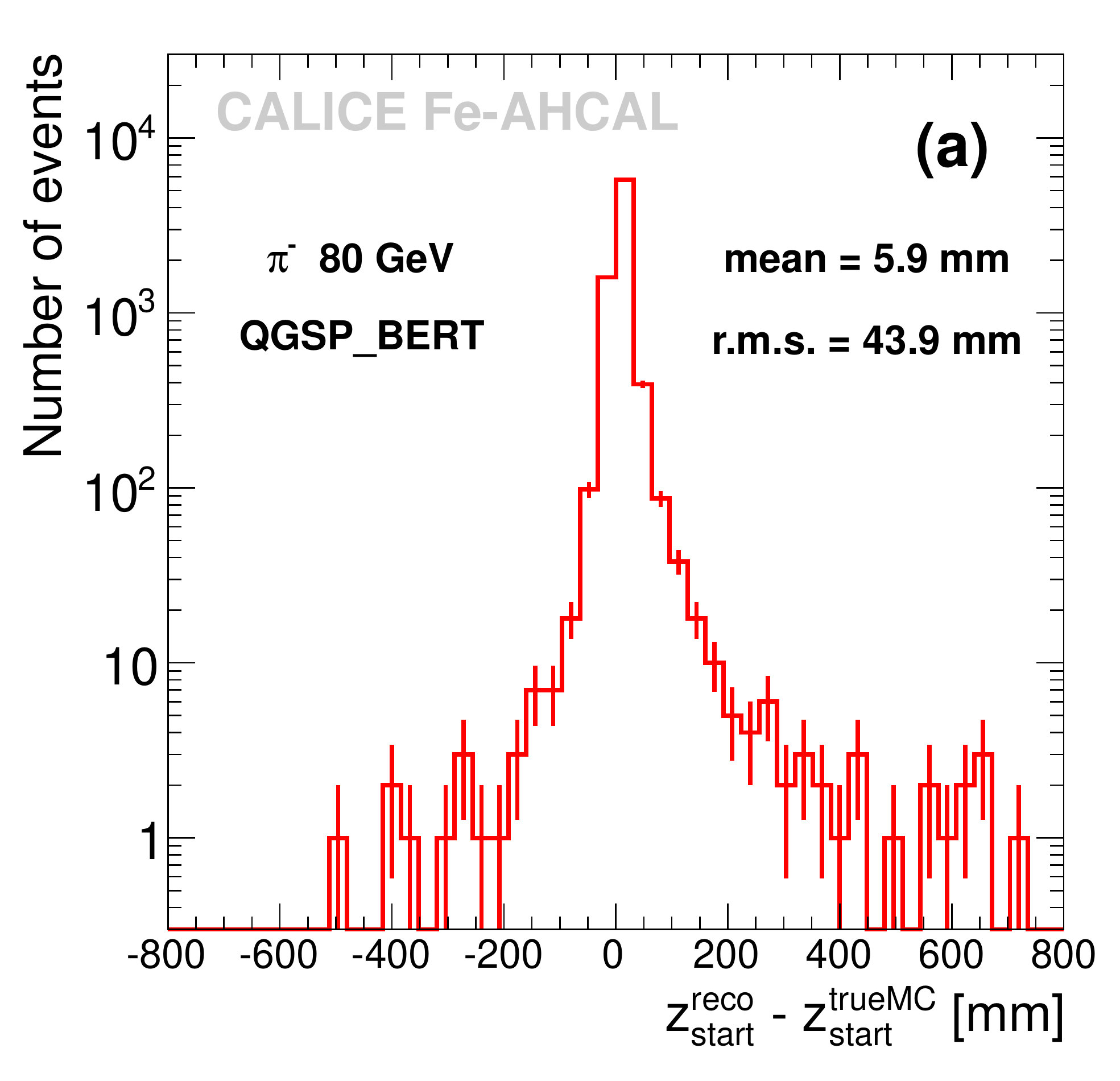}
 \includegraphics[width=7.5cm]{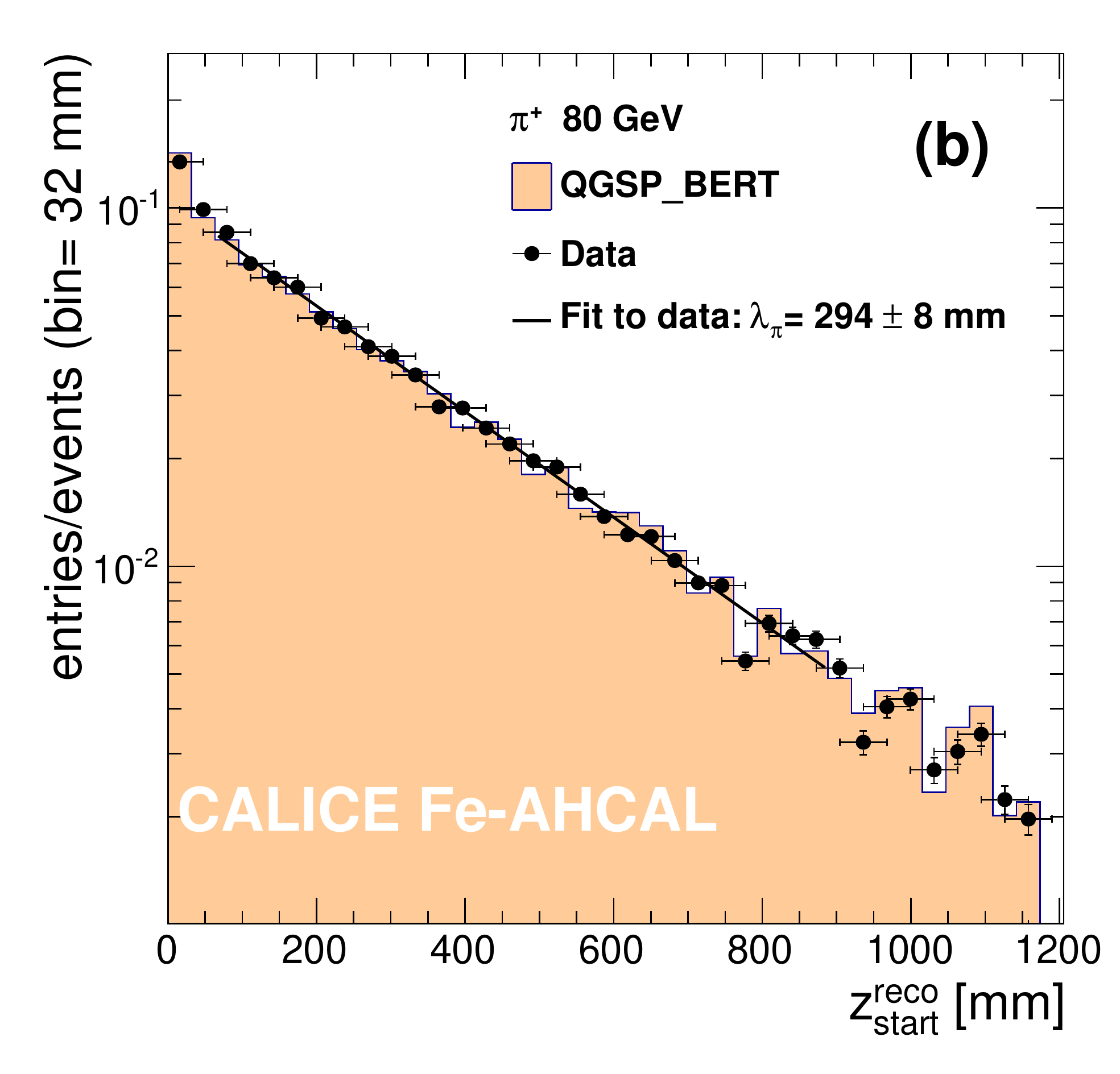}
 \caption{(a) Difference between the reconstructed shower start position and the true shower start position for simulated samples of 80~GeV pions. (b) Distributions of the reconstructed shower start position for data (circles) and simulations (filled histogram) for 80~GeV pions. The estimated nuclear interaction length obtained from the fit to data is shown in the legend. See text about fit details.}
 \label{fig:lamDist}
\end{figure}

The difference between the reconstructed shower start position, $z^{\mathrm{reco}}_{\mathrm{start}}$, and the real shower start position, $z^{\mathrm{trueMC}}_{\mathrm{start}}$, is shown in Fig.~\ref{fig:lamDist}(a) for 80~GeV pions. The values of $z^{\mathrm{trueMC}}_{\mathrm{start}}$ were obtained from the  {\sc Geant4} information about the position of the primary inelastic interaction of the incoming particle. The uncertainty of the algorithm is approximately $\pm$1 Fe-AHCAL layer ($\approx$32~mm). The algorithm finds the shower start within $\pm$1 layer of the true layer for $\sim$80\% of hadron events in the energy range studied, as determined with simulations.

Typical distributions of the reconstructed shower start position $z^{\mathrm{reco}}_{\mathrm{start}}$ with respect to the calorimeter front face are shown in Fig.~\ref{fig:lamDist}(b) for data and {\small FTFP\_BERT} physics list.  The distributions of the shower start layer for pions $f(z_{\mathrm{start}})$ can be fit with the exponential function

\begin{equation}
f(z_{\mathrm{start}}) = A \exp \left(-\frac{z_{\mathrm{start}}}{\lambda_{\pi}}\right),
\label{eq:lambdaPi}
\end{equation}

\noindent where $A$ is a normalisation factor and $\lambda_{\pi}$ is the estimated nuclear interaction length. The two first layers with large uncertainty for the shower start finding algorithm as well as several last layers were excluded from the fit, and the fit interval is from 65~mm to 900~mm for all samples.

The proton data samples contain pion contamination that varies from 5\% to 35\%. The distribution of the shower start position for a mixed sample of hadrons with different inelastic cross sections can be considered as a sum of two independent contributions 

\begin{equation}
f_{\mathrm{mix}}(z_{\mathrm{start}}) = A\:\left(\eta\:\exp\left(-\frac{z_{\mathrm{start}}}{\lambda_{\mathrm{p}}}\right) 
 + (1 - \eta)\:\exp\left(-\frac{z_{\mathrm{start}}}{\lambda_{\pi}}\right) \: \frac{\lambda_{\mathrm{p}}}{\lambda_{\pi}}\right),
\label{eq:lambdaP}
\end{equation}  

\noindent where the normalisation factor, $A$, and the nuclear interaction length for protons, $\lambda_{\mathrm{p}}$, are estimated variables, while the purity of proton sample, $\eta$, and the nuclear interaction length for pions, $\lambda_{\pi}$, are taken as previously determined and known parameters. The purity $\eta$ is estimated as described in Appendix~\ref{app:ppur}. The values of $\lambda_{\pi}$ are extracted from the fit to the distributions obtained from the corresponding pion samples at the same energy. 

The following procedure was used to estimate the systematic uncertainty due to the uncertainties of $\lambda_{\pi}$ and purity. The set of parameter values was generated using Gaussian distributions with the variance corresponding to the uncertainty of a given parameter and the generated values were used in the fit to proton data. The r.m.s. of the obtained distribution of $\lambda_{\mathrm{p}}$ is taken as a systematic uncertainty. The contributions from both parameters are summed up in quadrature.

\section{Estimate of the sample purity}
\label{app:ppur}

The purity of the sample, $\eta$, is estimated using the efficiency, $\epsilon$, to select contaminating events. The efficiency $\epsilon$ in turn is determined from an independent procedure, which does not involve information from the \u{C}erenkov counter.

In the current study, the pion-proton separation is based on the information from the \u{C}erenkov detector. The independent calorimeter-based muon identification procedure, as introduced in Section \ref{sec:evsel}, can be used to measure the efficiency of the \u{C}erenkov counter. Assuming that the efficiency for pions $\epsilon_{\mathrm{pion}}$ is approximately the same as for muons $\epsilon_{\mathrm{muon}}$ the value of $\epsilon_{\mathrm{pion}}$ can be calculated as

\begin{equation}
\epsilon_{\mathrm{pion}} \approx \epsilon_{\mathrm{muon}} = \frac{N^{\mathrm{cher}}_{\mathrm{muon}}}{N^{\mathrm{total}}_{\mathrm{muon}}},
\label{eq:muEff}
\end{equation}  

\noindent where $N^{\mathrm{cher}}_{\mathrm{muon}}$ is the number of identified muons that gave a signal in the \u{C}erenkov detector, $N^{\mathrm{total}}_{\mathrm{muon}}$ is the total number of muons identified using the calorimeter-based procedure. 

The purity of the proton sample, $\eta_{\mathrm{p}}$, i.e. the ratio of the true number of protons to the number of identified protons, can be obtained as 

\begin{equation}
\eta_{\mathrm{p}} = 1 - \frac{N_{\pi}}{N_{\mathrm{p}}} \: \left (\frac{1 - \epsilon_{\mathrm{pion}}}{\epsilon_{\mathrm{pion}}}\right ),
\label{eq:prPur}
\end{equation}  

\noindent where  $N_{\pi}$ ($N_{\mathrm{p}}$) is the number of pions (protons) in the selected samples identified using the \u{C}erenkov counter. The uncertainties on $\epsilon_{\mathrm{muon}}$ are estimated from the available statistics of the muon event sample and are propagated to the uncertainty of $\eta_{\mathrm{p}}$. As the pressure in the gaseous \u{C}erenkov detector used was set well below the proton threshold, the misidentification of protons is negligible.

The efficiency of positron identification, $\epsilon_{\mathrm{pos}}$, is estimated from the dedicated positron runs taken without an electromagnetic calorimeter as

\begin{equation}
\epsilon_{\mathrm{pos}} = \frac{N_{\mathrm{selected}}}{N_{\mathrm{total}}},
\label{eq:posEff}
\end{equation}  

\noindent where $N_{\mathrm{selected}}$ is the number of selected positrons from the pure positron sample with $N_{\mathrm{total}}$ events. 
 Then the purity of the selected pion samples are calculated from the equation
 
\begin{equation}
\eta_{\pi} = 1 - \frac{N_{\mathrm{pos}}}{N_{\pi}} \: \left (\frac{1 - \epsilon_{\mathrm{pos}}}{\epsilon_{\mathrm{pos}}}\right ),
\label{eq:posPur}
\end{equation} 

\noindent where  $N_{\mathrm{pos}}$ ($N_{\pi}$) is the number of positrons (pions) in the selected samples, identified using the positron selection procedure, which is described in Section \ref{sec:evsel}.

\end{document}